\pdfoutput=1
\documentclass[
aps,
physrev,
superscriptaddress,
citeautoscript,
amsmath,
amssymb,
amsfonts,
showkeys,
floatfix,
preprint,
]{revtex4-2}

\usepackage{xparse}
\usepackage{ifthen}

\usepackage{xcolor}
\usepackage{graphicx}

\usepackage{comment}

\usepackage{dcolumn}

\usepackage{amssymb}

\usepackage{natbib}
\usepackage{siunitx}

\usepackage{hyperref}
\hypersetup{
  plainpages=false,
  unicode=true,
  colorlinks=true,
  linkcolor=blue,
  citecolor=blue,
  urlcolor=blue,
}
\usepackage[all]{hypcap}

\DeclareSIUnit\angstrom{\text{Å}}

\newboolean{draft}
    \setboolean{draft}{true}
\NewDocumentCommand\todo{m}%
{%
  {\color{blue} (TODO: #1)}%
}
\NewDocumentCommand\change{om}%
{%
  \ifthenelse{\boolean{draft}}{
    \IfNoValueTF{#1}{}%
    {%
      {\color{gray}[#1]}
    }%
    {\color{orange}#2}%
  }%
  {#2}%
}

\NewDocumentCommand{\wf}{O{i}}{\ket{\Phi_{#1}}}
\NewDocumentCommand{\vwf}{s O{i} O{n} O{} O{}}{
  \chi_{
    #3}^{#4}
  \IfBooleanTF#1
  {}
  {(\Q_{#5})}
}

\newcommand{\ket}[1]{
  \mathchoice%
  {\left|#1\right\rangle}       
  {|#1\rangle}                  
  {|#1\rangle}                  
  {|#1\rangle}                  
}
\newcommand{\braket}[2]{\left\langle#1\middle|#2\right\rangle}

\begin{document}


\title{Theory of the divacancy in 4H-SiC: Impact of Jahn--Teller effect on optical properties}

\author{Vytautas Žalandauskas}
\email{vytautas.zalandauskas@ftmc.lt}
\affiliation{Center for Physical Sciences and Technology (FTMC), Vilnius LT--10257, Lithuania}
\author{Rokas Silkinis}
\affiliation{Center for Physical Sciences and Technology (FTMC), Vilnius LT--10257, Lithuania}
\author{Lasse Vines}
\affiliation{Department of Physics/Centre for Materials Science and Nanotechnology, University of Oslo, P.O. Box 1048, Blindern, Oslo N-0316, Norway}
\author{Lukas Razinkovas}
\affiliation{Center for Physical Sciences and Technology (FTMC), Vilnius LT--10257, Lithuania}
\affiliation{Department of Physics/Centre for Materials Science and Nanotechnology, University of Oslo, P.O. Box 1048, Blindern, Oslo N-0316, Norway}
\author{Marianne Etzelmüller Bathen}
\email{m.e.bathen@fys.uio.no}
\affiliation{Department of Physics/Centre for Materials Science and Nanotechnology, University of Oslo, P.O. Box 1048, Blindern, Oslo N-0316, Norway}




\begin{abstract}
Understanding the optical properties of color centers in silicon carbide is essential for their use in quantum technologies, such as single-photon emission and spin-based qubits. In this work, first-principles calculations were employed using the r$^{2}$SCAN density functional to investigate the electronic and vibrational properties of neutral divacancy configurations in 4H-SiC. Our approach addresses the dynamical Jahn--Teller effect in the excited states of axial divacancies. By explicitly solving the multimode dynamical Jahn--Teller problem, we compute emission and absorption lineshapes for axial divacancy configurations, providing insights into the complex interplay between electronic and vibrational degrees of freedom. The results show strong alignment with experimental data, underscoring the predictive power of the methodologies. Our calculations predict spontaneous symmetry breaking due to the pseudo Jahn--Teller effect in the excited state of the $kh$ divacancy, accompanied by the lowest electron--phonon coupling among the four configurations and distinct polarizability. These unique properties facilitate its selective excitation, setting it apart from other divacancy configurations, and highlight its potential utility in quantum technology applications. These findings underscore the critical role of electron--phonon interactions and optical properties in spin defects with pronounced Jahn--Teller effects, offering valuable insights for the design and integration of quantum emitters for quantum technologies.
\end{abstract}

\maketitle


\section{Introduction}

Silicon carbide (SiC) is a wide band gap material with great potential for high-power and high-frequency electronic devices~\cite{She_2017} and quantum technologies (QT). Deep-level defects in the SiC crystal structure can significantly affect carrier mobility and lifetime, negatively impacting the performance of power devices~\cite{Kimoto_2014,Klein_2006,Ghezellou_2023}. While not all deep-level defects are optically active, some can introduce discrete energy levels deep within the band gap that can be optically addressed. At sufficiently low densities, such \textit{color centers} can serve as sources of non-classical light~\cite{Aharonovich_2016}. Furthermore, some states reveal a paramagnetic spin structure with long coherence times even at room temperature (RT), and may therefore be optically accessible for spin initialization and readout protocols~\cite{Castelletto_2020}. These properties position solid-state color centers as promising candidates for single-photon emitters (SPEs) and spin qubits. When integrated with advanced material processing and fabrication capabilities, color centers in SiC emerge as a robust platform for future solid-state quantum technologies~\cite{Weber_2010,Awschalom_2018}.

SiC hosts a variety of defects with QT-compatible properties~\cite{Zhang_2020,Son_2020,Bathen_QuTe_2021}. Key intrinsic defects such as the silicon vacancy (V$_{\text{Si}}$), divacancy (V$_{\text{Si}}$V$_{\text{C}}$), and carbon antisite-vacancy pair (C$_{\text{Si}}$V$_{\text{C}}$) are introduced by irradiation and annealing and are proven to be photostable SPEs at RT ~\cite{Widmann_2014,Castelletto_2014,Christle_2015}. C$_{\text{Si}}$V$_{\text{C}}$ emits in the visible spectrum (640--680~nm), V$_{\text{Si}}$ in the near-infrared (858--916~nm)~\cite{Castelletto_2014,Widmann_2014}, whereas the divacancy exhibits zero-phonon line (ZPL) energies closer to the telecommunications band at 1040--1140~nm~\cite{ Christle_2015,Christle_2017}. A  deterministic single-photon emission in the infrared, such as that from the divacancy, is essential for fiber optic technology, enabling low-loss, secure long-distance quantum communication and cryptography. Furthermore, the divacancy's electron spins show outstanding longevity, with coherence times exceeding those of nitrogen-vacancy (NV) centers in diamond, with V$_{\text{Si}}$V$_{\text{C}}$ exhibiting $T_2>5$~s~\cite{Anderson_2022}. Thus, the divacancy stands out as a versatile candidate for QT applications, functioning as both an SPE and a spin qubit.

Although color centers in crystals offer numerous advantages as SPEs and qubits, including RT operation, photostability, and scalability, the solid-state matrix also presents several challenges. These include both inhomogeneous and homogeneous optical broadening effects. Inhomogeneous broadening can be mitigated by improving the material quality. However, homogeneous broadening, which results from interactions with the continuum of vibrational degrees of freedom in solids, is more challenging to control. Such electron--phonon interactions not only lead to the emergence of phonon sidebands that reduce the number of photons emitted at the ZPL but also contribute to its temperature broadening, thereby diminishing the photon indistinguishability properties of the emitter. Furthermore, spin--phonon coupling plays a crucial role in determining the coherence properties of spin systems, as well as in driving the mechanisms that lead to intersystem crossing, which are essential for spin initialization and optical-spin contrast in optically detected magnetic resonance (ODMR) experiments. The theoretical description of the vibrational structure and electron--phonon coupling mechanisms of deep-level defects poses significant challenges, as defects exhibit characteristics of both atomic and solid-state physics, necessitating advanced methodologies that extend beyond conventional \textit{ab initio} calculations~\cite{Alkauskas_2014,Razinkovas_2021}.

Advancements in understanding crystal defects as potential quantum systems have stimulated a renewed focus on the Jahn--Teller (JT) effect in color centers~\cite{Abtew_2011,Thiering_2017,Zhang_2018,Thiering_2018,Thiering_2019,Thiering_2021,Carbery_2024}. This effect emerges when degenerate or near-degenerate electronic states interact with symmetry-breaking vibrational modes. This coupling intricately mixes the electronic and vibrational degrees of freedom, giving rise to complex dynamics that influence the system's electronic and optical properties~\cite{Ham_1965,Longuet_1958,Bersuker_2006}. Specifically, the JT interaction induces unique vibronic states, distinct from their adiabatic counterparts, manifesting as characteristic features in the absorption and emission spectra. These effects underscore the pivotal role of electron--phonon coupling in shaping the optical signatures of these systems. The description of JT systems typically employs a phenomenological model, approximating the numerous JT active modes with a single effective degenerate mode~\cite{Longuet_1958,Abtew_2011,Thiering_2017,Thiering_2018}. However, accurately capturing optical signatures often necessitates solving the multi-mode JT problem, as recently demonstrated in modeling optical spectra for the negatively charged nitrogen-vacancy~\cite{Razinkovas_2021} and nickel-vacancy~\cite{Silkinis_2024} centers in diamond.

In this work, first-principles calculations were performed utilizing the recently developed r$^{2}$SCAN density functional~\cite{Furness_2020} to investigate the electronic and vibrational properties of the four distinct configurations of the neutral divacancy in the 4H polytype of SiC. Previous work~\cite{Jin_2021} modeled emission lineshapes for three divacancy configurations but did not explicitly solve the multimode dynamic Jahn--Teller (DJT) problem present in the excited state of axial divacancies or calculate absorption lineshapes. This work addresses these gaps by providing calculated emission and absorption lineshapes for all four divacancy configurations and explicitly treating the multimode DJT problem for axial divacancies, which exhibit a pronounced DJT effect. The dynamical coupling between electronic and vibrational degrees of freedom can create distinct vibronic states, resulting in specific emission or absorption spectra features. Herein, we demonstrate that this must be considered to correctly capture the optical properties of the divacancy in 4H-SiC and discuss the implications of these findings for other systems.


\section{Results}

\subsection{Electronic structure}

\begin{figure}[t]
  \includegraphics[width=0.30\textwidth]{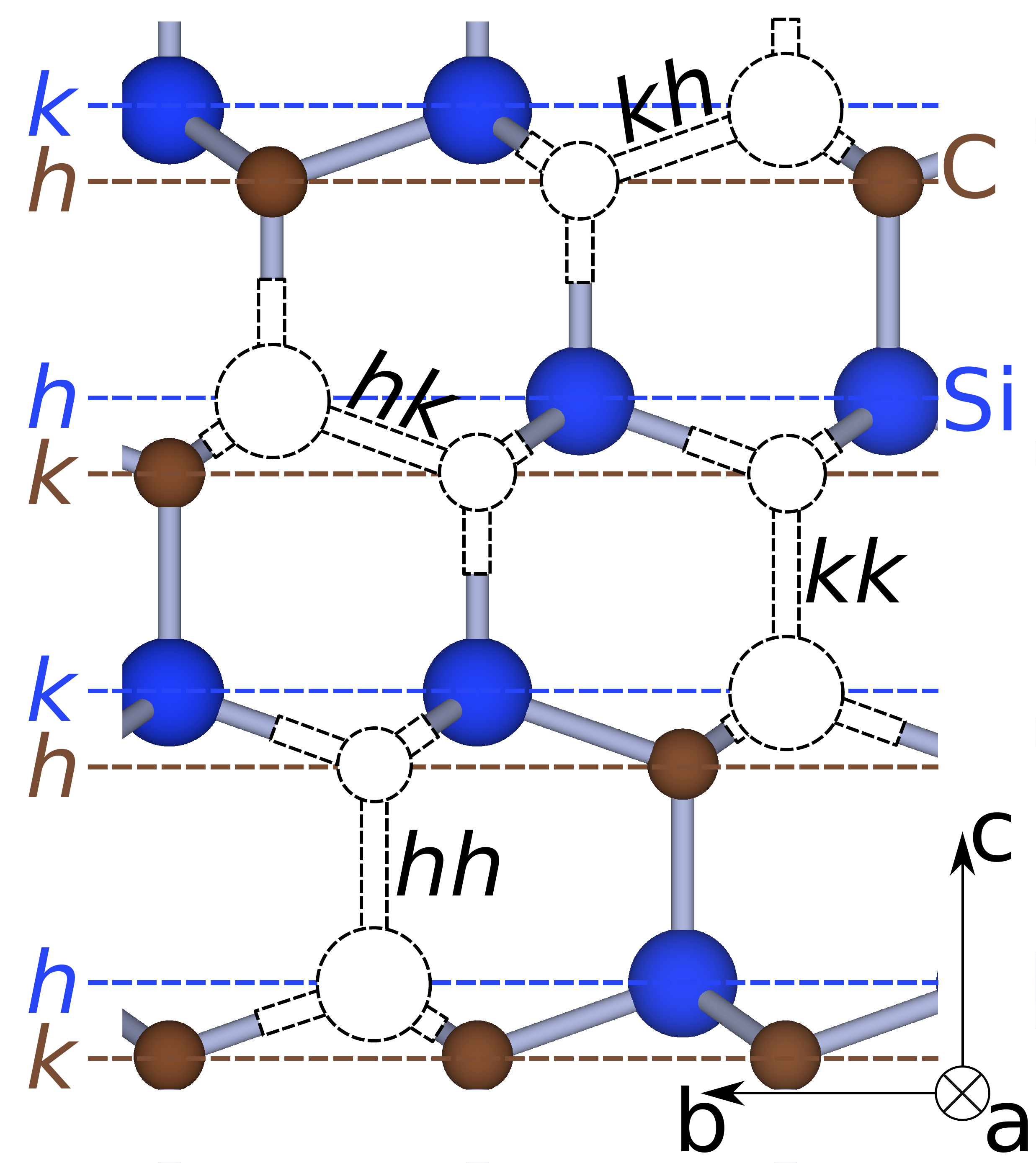}
  \caption{\textbf{Structure of neutral divacancy (V$_{\text{Si}}$V$_{\text{C}}^{0}$ or VV$^{0}$) defects in 4H-SiC.} Ball and stick representation of vacancy pairs depicted as hollow spheres. The carbon and silicon atoms are represented by brown and blue spheres, respectively. Four nonequivalent configurations ($hh$, $kk$, $hk$, $kh$) of divacancy defects are shown. The planes labeled \textit{h} and \textit{k} indicate the symmetry of lattice sites.\label{fig:defects}}
\end{figure}

Divacancies in the neutral charge state were studied by performing calculations within the spin-polarized density functional theory (DFT) framework and using the r$^2$SCAN meta-GGA functional~\cite{Furness_2020}. In the 4H-SiC polytype, each atomic site can take on either a hexagonal (\textit{h}) or pseudo-cubic (\textit{k}) configuration. Thus, four distinct defect configurations must be considered for a binary complex, such as the silicon-carbon divacancy that can take on different symmetries and exhibit different electro-optical properties (see Fig.~\ref{fig:defects}).

\begin{figure*}
  \centering
  \includegraphics[width=\textwidth]{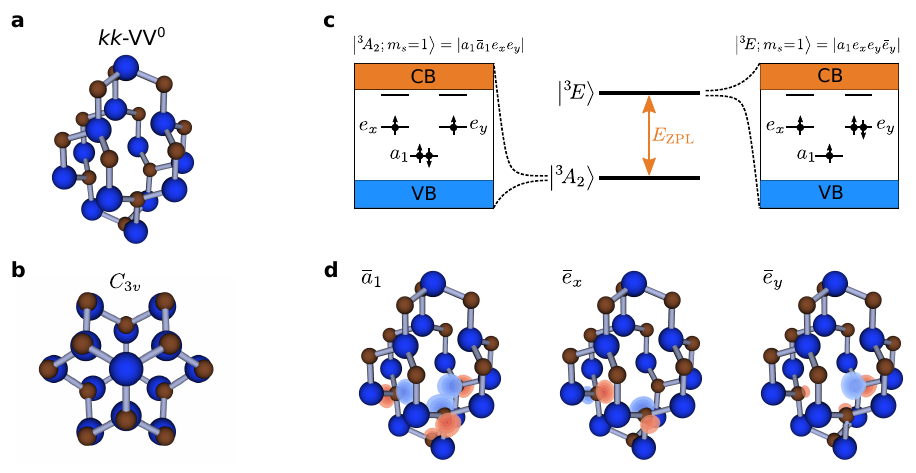}
  \caption{\textbf{Description of electronic structure of axial divacancies.} \textbf{a} Ball and stick representation of the $kk$-VV$^{0}$ defect in 4H-SiC. \textbf{b} Top-down view of $kk$-VV$^{0}$ along the $c$-axis showcasing the $C_{3v}$ symmetry of the defect. \textbf{c} Schematic representation of single-particle defect levels in the band gap of 4H-SiC showing the occupancy of these levels in the ground $^{3}A_{2}$ state and excited $^{3}E$ state of $kk$-VV$^{0}$. \textbf{d} Isosurfaces of the real part of the single-particle orbitals associated with the defect levels $\overline{a}_{1}$, $\overline{e}_{x}$ and $\overline{e}_{y}$. The color (blue/red) represents the sign ($+/-$) of the orbital.}
  \label{fig:MO}
\end{figure*}

The electronic structure of neutral divacancies in 4H-SiC can be described using three molecular orbitals derived from the carbon dangling bonds, as depicted in Fig.~\ref{fig:MO}c~and~\ref{fig:MO}d. Axial divacancies, $hh$-VV$^{0}$ and $kk$-VV$^{0}$, which exhibit $C_{3v}$ point-group symmetry (see Fig.~\ref{fig:MO}a and \ref{fig:MO}b), consist of an orbital of the $a_{1}$ irreducible representation and a doublet of orbitals with $e$-symmetry (see Fig.~\ref{fig:MO}c). DFT calculations reveal an additional orbital doublet of $e$-symmetry within the band gap close to the conduction band edge, originating from silicon dangling bonds; these orbitals, however, remain optically inactive under standard excitation conditions. The ground state is a spin-triplet with $A_{2}$ orbital symmetry, designated as $^{3}\!A_{2}$ for axial divacancies, and is characterized by the orbital configuration $a_{1}^{2}e^{2}$. This state maintains a single-determinant wave function $|a_{1}\bar{a}_{1}e_{x}e_{y}|$ for the spin projection $m_{s}=1$. The transition of a spin-down electron from the $a_{1}$ an $e$ orbital located on carbon, leads to the excited spin-triplet state $^{3}E$ which is characterized by a degenerate orbital configuration $a_{1}e^{3}$. For $m_{s}=1$, this electron configuration manifests as two single-determinant wave functions $|a_{1}e_{x}e_{y}\bar{e}_{y}|$ and $|a_{1}\bar{e}_{x}e_{x}e_{y}|$. The symmetry dictates that the dipole moment for the transition ${^{3}A_{2}} \leftrightarrow {^{3}E}$ in 4H-SiC is polarized within the $c$-plane.

In contrast to axial divacancies, $hk$ and $kh$ divacancies exhibit reduced $C_{1h}$ point group symmetry in the ground state, impacting their electronic and optical properties. The $hk$ divacancy displays three lowest-lying levels within the band gap, ordered from lowest to highest energy respectively as $a'(1)$, $a''$, and $a'(2)$, where $a'$ and $a''$ denote irreducible representations of the $C_{1h}$ point group. DFT calculations predict that the ground state of the $kh$ divacancy has orbital symmetries denoted as $a'(1)$, $a'(2)$, and $a''$ while the excited state of this divacancy has a reduced $C_{1}$ symmetry with its orbitals all having $a$-symmetry. Illustrations of single-particle levels for each basal configuration are provided in Supplementary Fig. 1. Using group-theoretical analysis based on the $C_{1h}$ point group, different polarization of the ZPL can be expected for $hk$ and $kh$ divacancies. In the $hk$ divacancy, the optical transition occurs between the $a'(1)$ and $a''$ levels, involving a change in parity from $a'$ to $a''$. Consequently, the transition dipole moment (TDM) must transform according to the $a''$ irreducible representation. This symmetry implies that the polarization of the emitted or absorbed light will be oriented perpendicular to the mirror plane $\sigma_h$, resulting in a dipole moment polarized within the $c$-plane (similar to $hh$-VV$^{0}$ and $kk$-VV$^{0}$). For the $kh$ divacancy, however, the optical transition occurs between the $a'(1)$ and $a'(2)$ levels. Since there is no change in parity, the polarization of the emitted or absorbed light is within the symmetry reflection plane of the defect, with most of its projection perpendicular to the $c$-plane (theoretical transition dipole matrix elements are provided in Supplementary Table 1). This analysis is consistent with previously reported experimental and theoretical findings~\cite{Davidsson_2020,Stenlund_2023,Shafizadeh_2024}.

\begin{table}
  \caption{\label{tab:ZPL} Calculated zero-phonon line (ZPL) energies (in units of eV) and zero-field splitting (ZFS) parameters (in units of GHz) compared to experimental values.}
  \begin{ruledtabular}
    \begin{tabular}{l|cc|cc}
      Configuration & Calc. ZPL & Exp. ZPL \cite{Falk_2013} & Calc. ZFS & Expt. ZFS \cite{Falk_2013} \\
      \hline
      \textit{hh} & 1.079 & 1.095 & 1.335 & 1.336 \\
      \textit{kk} & 1.081 & 1.096 & 1.285 & 1.305 \\
      \textit{kh} & 1.062 & 1.119 & 1.292 & 1.222 \\
      \textit{hk} & 1.100 & 1.150 & 1.328 & 1.334 \\
    \end{tabular}
  \end{ruledtabular}
\end{table}

The calculated ZPL energies and zero-field splitting (ZFS) parameters for all divacancy configurations are compared with the experimental results and presented in Table~\ref{tab:ZPL}. The results obtained using the r$^{2}$SCAN functional are in excellent agreement with experimental data~\cite{Falk_2013} (see Supplementary Table 2 and Supplementary Table 3 for a comparison of results obtained using the SCAN~\cite{Sun_2015} functional and other theoretical works). From the mean absolute errors, SCAN and r$^2$SCAN functionals enhance the accuracy compared to other calculations and come closest to the experimental values while requiring only a fraction of the computational cost needed for hybrid functionals.

\subsection{Jahn--Teller effect}

\begin{figure*}
  \centering
  \includegraphics[width=\textwidth]{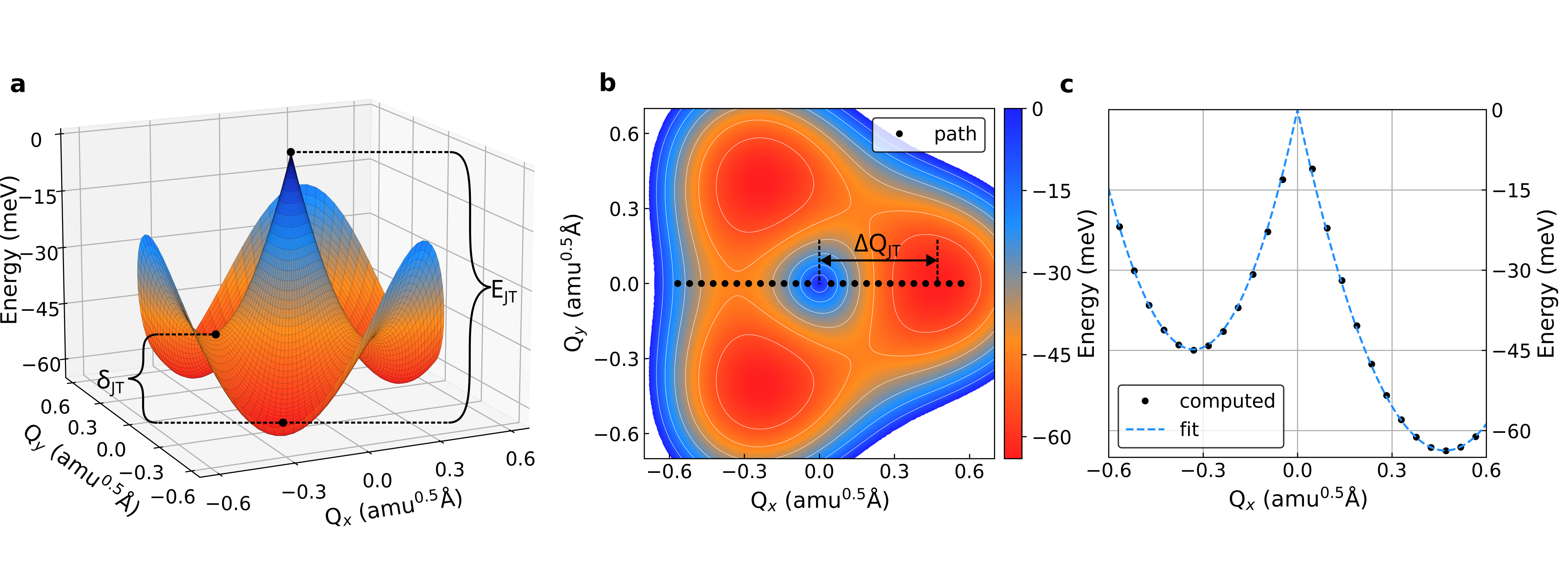}
  \caption{\textbf{Adiabatic potential energy surface (APES) of the \textit{kk}-VV$^{0}$ excited state $\mathbf{^{3}\textit{E}}$}. \textbf{a} APES of the lower branch of the $^{3}E$ state, which takes a "tricorn sombrero" shape with three minima and three saddle points. $Q_{x}$, $Q_{y}$ configuration coordinates represent the collective motion of effective $e$-symmetry modes. $E_{\text{JT}}$ is the JT stabilization energy, which is the difference in energy between the high-symmetry and distorted configurations. The three minima are separated by three barriers with the energy of $\delta_{\text{JT}}$. \textbf{b} Contour plot of the APES. \textbf{c} Computed potential energy curve along the path defined in \textbf{b}. The potential energy curve is fitted by Eq. (\ref{eqn:JT}) (see text).}
  \label{fig:JT-APES}
\end{figure*}

The excited state $^{3}E$ of the axial $hh$ and $kk$ divacancies is a degenerate orbital doublet with orbital configurations $a_{1}e_{x}^{1}e_{y}^{2}$ and $a_{1}e_{x}^{2}e_{y}^{1}$. This degeneracy results in an intricate non-adiabatic coupling between electronic $E$ states and $e$-symmetry degenerate vibrational modes forming an $E\otimes(e\oplus e \cdots)$ JT system. Since the coupling is not too strong, it leads to a DJT effect with dynamical distortions along symmetry-breaking directions, generating a distinct array of vibronic states and giving rise to specific features within optical spectra. The electronic structure of this DJT system is analogous to that of the negatively charged NV center in diamond, where the interplay between electronic and vibrational states due to the DJT effect has been well-studied~\cite{Fu_2009,Abtew_2011,Thiering_2017,Razinkovas_2021}.

To analyze the adiabatic potential energy surface (APES) of the excited triplet state $^{3}E$, an effective single-degenerate-mode model for quadratic \mbox{$E \otimes e$} vibronic coupling is used to measure the effective parameters describing a two-dimensional representation of a multi-dimensional potential energy surface~\cite{Bersuker_2006}. The expression for the APES in the space of polar mass-weighted displacement coordinates $Q_{x}$ and $Q_{y}$ (where $Q_{x}$ and $Q_{y}$ describe a pair of effective coordinates that parametrize a collective ionic motion along the $e$-symmetry direction) takes the form
\begin{equation}
  \label{eqn:JT}
  \epsilon (\rho, \phi) = \frac{1}{2} \kappa \rho^{2} \pm \rho [F^{2} + G^{2} \rho^{2} + 2 F G \rho \cos(3\phi)]^{1/2},
\end{equation}
where $\rho = \sqrt{(Q_{x}^{2}+Q_{y}^{2})}$, $\phi = \arctan(Q_{y}/Q_{x})$, $\kappa$ is the elastic force constant, $F$ is the linear vibronic constant, and $G$ is the quadratic vibronic constant. The values of the three vibronic constants, mass-weighted displacements, and details of first-principles calculation of the APES are provided in Supplementary Note 4. The lower branch of the $^{3}E$ APES takes a "tricorn sombrero" shape with three minima and three saddle points (see Fig.~\ref{fig:JT-APES}a). The three minima have a JT stabilization energy $E_{\text{JT}}$ of 69~meV for $hh$-VV$^{0}$ and 64~meV for $kk$-VV$^{0}$. These three minima are separated by energy barriers $\delta_{\text{JT}}$ of 23~meV for $hh$-VV$^{0}$ and 19~meV for $kk$-VV$^{0}$. These results are in good agreement with other theoretical studies, where the use of the hybrid HSE06 functional yielded JT stabilization energy and barrier values for the $hh$-VV$^{0}$ defect of 74 meV and 18 meV, respectively~\cite{Bian_2024}. Fig.~\ref{fig:JT-APES}b shows the APES contour plot and the path used to calculate the potential energy curve (see Fig.~\ref{fig:JT-APES}c) and parametrize the JT relaxation. For the analysis of the multi-mode vibronic states of axial divacancy, including the quadratic JT effect would significantly increase the complexity of an already intricate problem. Therefore, we use linear JT theory to analyze the impact of the JT effect on vibronic dynamics and its influence on the optical properties of axial divacancies.

For basal divacancies, the excited states exhibit non-degenerate energy levels that are closely spaced. $\Delta$SCF calculations indicate that in the ground-state geometry, the energy separation between these excited states is $62~\mathrm{meV}$ for the $hk$ divacancy and $100~\mathrm{meV}$ for the $kh$ divacancy. This small distance suggests the potential presence of a pseudo-Jahn--Teller (PJT) effect, where symmetry-breaking vibrational modes of $a''$ symmetry couple to electronic states that are close in energy. In static adiabatic calculations, the PJT effect modifies the harmonic potential. For weak coupling, this manifests as changes in the effective frequency of PJT-active modes, while stronger coupling can lead to the appearance of double minima and symmetry breaking~\cite{Bersuker_2006}. Our results reveal that the $kh$ divacancy in the excited state undergoes symmetry breaking, providing additional stabilization energy of $15~\mathrm{meV}$. In contrast, no such symmetry breaking is observed for the $hk$ divacancy, indicating weaker PJT coupling. A comprehensive analysis of the PJT effect necessitates the explicit determination of electron--phonon coupling parameters ($\langle \psi_{i} | \partial \psi_{i} / \partial Q_{k} \rangle$) directly from the wave functions. In the case of pure degeneracy, these parameters can be inferred from the shape of the APES. However, such an analysis is beyond the scope of the present study. As a result, for the examination of the optical properties of basal divacancies, we neglect the non-adiabatic nature of the PJT effect and instead rely on the Huang--Rhys (HR) theory~\cite{Huang_1950}. This approximation could be justified for the emission process~\cite{Razinkovas_2021}, where the final state can be effectively treated as adiabatic. Nonetheless, a more detailed investigation is necessary to rigorously address the absorption spectra, which is left for future work.

\subsection{Vibrational properties}

\begin{figure*}
  \centering
  \includegraphics[width=\textwidth]{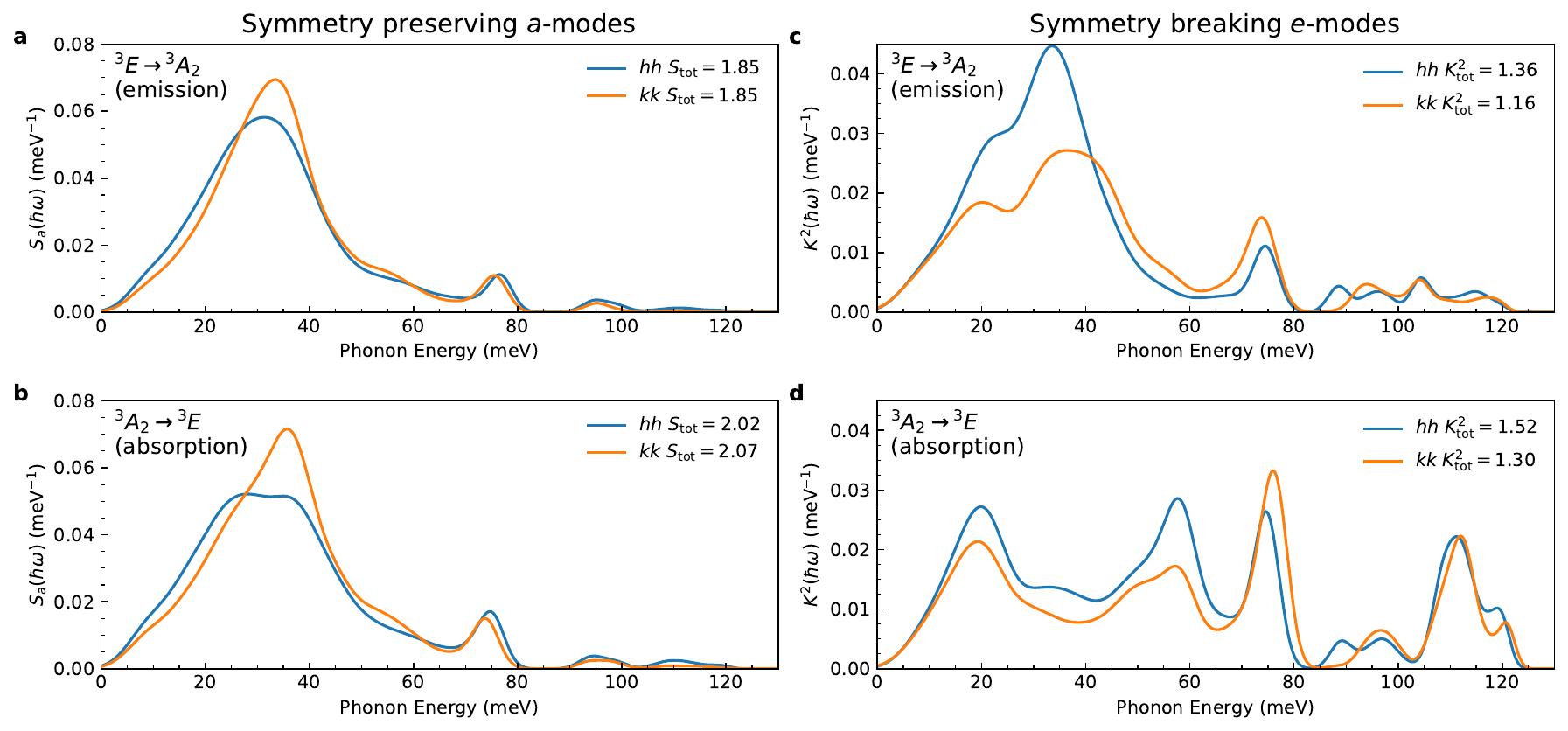}
  \caption{\textbf{Spectral densities of electron--phonon coupling for axial divacancies.} \textbf{a} Spectral densities $S_{a_{1}} (\hbar\omega)$ associated with $a_{1}$-symmetry phonons of the ${^3E}\to{^3A_{2}}$ and \textbf{b} ${^3A_{2}}\to{^3E}$ transitions. \textbf{c} Spectral densities of JT linear coupling $K^{2} (\hbar\omega)$ associated with $e$-symmetry phonons of the ${^3E}\to{^3A_{2}}$ and \textbf{d} ${^3A_{2}}\to{^3E}$ transitions. The spectral densities are for $hh$ and $kk$ divacancies. The total HR factor $S_{\text{tot}}$ and the total JT coupling $K^{2}_\mathrm{tot}$ are included in the panels. Results are for the $(23\times23\times7)$ supercell with 29\,622 atomic sites.}
  \label{fig:ep-coupling-axial}
\end{figure*}

Vibrational structures of the ground and excited states and the subsequent relaxation profiles were calculated to obtain electron--phonon coupling densities for all four divacancy configurations. The embedding methodology of Refs.~\cite{Alkauskas_2014,Razinkovas_2021} was employed to simulate these parameters in supercells containing nearly 30\,000 atomic sites, approaching the dilute limit.

The calculated spectral densities of electron--phonon coupling for the emission (${^3E} \to {^3A_{2}}$) and absorption (${^3A_{2}} \to {^3E}$) processes for the axial divacancies are presented in Fig.~\ref{fig:ep-coupling-axial}. The coupling to JT inactive ($a_{1}$-symmetry) modes is represented by $S_{a_{1}}(\hbar\omega)=\sum_{k}S_{k}\delta(\hbar\omega-\hbar\omega_{k})$ (see Fig.~\ref{fig:ep-coupling-axial}a and \ref{fig:ep-coupling-axial}b), where $S_{k}$ is the HR factor~\cite{Huang_1950,Alkauskas_2014,Razinkovas_2021} for vibrational mode $k$, quantifying the average number of $a_{1}$-symmetry phonons emitted during an optical transition. The linear JT coupling spectral density for $e$-symmetry modes is given by $K^{2}(\hbar\omega)=\sum_{k}K^{2}_{k}\delta(\hbar\omega-\hbar\omega_{k})$ (see Fig.~\ref{fig:ep-coupling-axial}c and \ref{fig:ep-coupling-axial}d), where $K_{k}$ is the dimensionless vibronic coupling constant~\cite{Obrien_1980,Razinkovas_2021,Silkinis_2024} quantifying the coupling between the $e$-symmetry vibrational doublet $k$ and components of the degenerate electronic state $E$. The total HR factor $S_\mathrm{tot}=\sum_{k}S_{k}$ and the total JT coupling $K^{2}_\mathrm{tot}=\sum_{k}K^{2}_{k}$ are included in Fig.~\ref{fig:ep-coupling-axial}.

The parameters $S_k$ and $K^2_k$ describe alterations in the APES and are directly related to relaxation energies. The relaxation energy along the symmetry-preserving direction following the vertical transition $A \to E$ is given by $\Delta E_{a_{1}} = \sum_k \hbar \omega_k S_k$. In contrast, the linear theory defines the JT stabilization energy as $\Delta E_{\text{JT}} = \sum_k \hbar \omega_k K^2_k$. Our computations yield $\Delta E_{a_{1}} = 74.2$~meV for $hh$-VV$^{0}$ and 76.1~meV for $kk$-VV$^{0}$. This suggests that the contribution to the electron--phonon coupling from JT-active modes ($E_{\text{JT}} = 69$~meV for $hh$ and 64~meV for $kk$) is comparable to that from symmetry-preserving modes.

During the ${^3E} \to {^3A_{2}}$ optical transition, the total HR factor of $a_{1}$-symmetry modes is found to be equal for $hh$ and $kk$ divacancies, although the $hh$ divacancy has a higher total JT coupling ($K^{2} = 1.36$) than the $kk$ divacancy ($K^{2} = 1.16$). The main difference in the vibrational structures between emission and absorption processes lies in the contribution from $e$-symmetry phonons. The JT linear coupling $K^{2}(\hbar\omega)$ for the absorption process is overall shifted to higher energy levels compared to the emission process, particularly for phonon energies above 100~meV. In the case of the ${^3A_{2}}\to{^3E}$ optical transition, the total calculated JT coupling of $e$-symmetry phonons is 1.52 ($hh$-VV$^{0}$) and 1.30 ($kk$-VV$^{0}$), respectively, which is several times stronger than that for the NV$^{-}$ center in diamond~\cite{Razinkovas_2021}. This warrants the use of multimode DJT theory to model the absorption lineshapes for the axial divacancies. During the emission process, degeneracy resides in the initial state, and therefore, HR theory can be utilized to calculate the emission lineshapes~\cite{Razinkovas_2021}. For a complete comparison, the DJT theory was still applied to calculate the spectral densities of JT linear coupling and subsequent luminescence lineshapes for axial divacancies.

\begin{figure}
  \centering
  \includegraphics[width=0.95\textwidth]{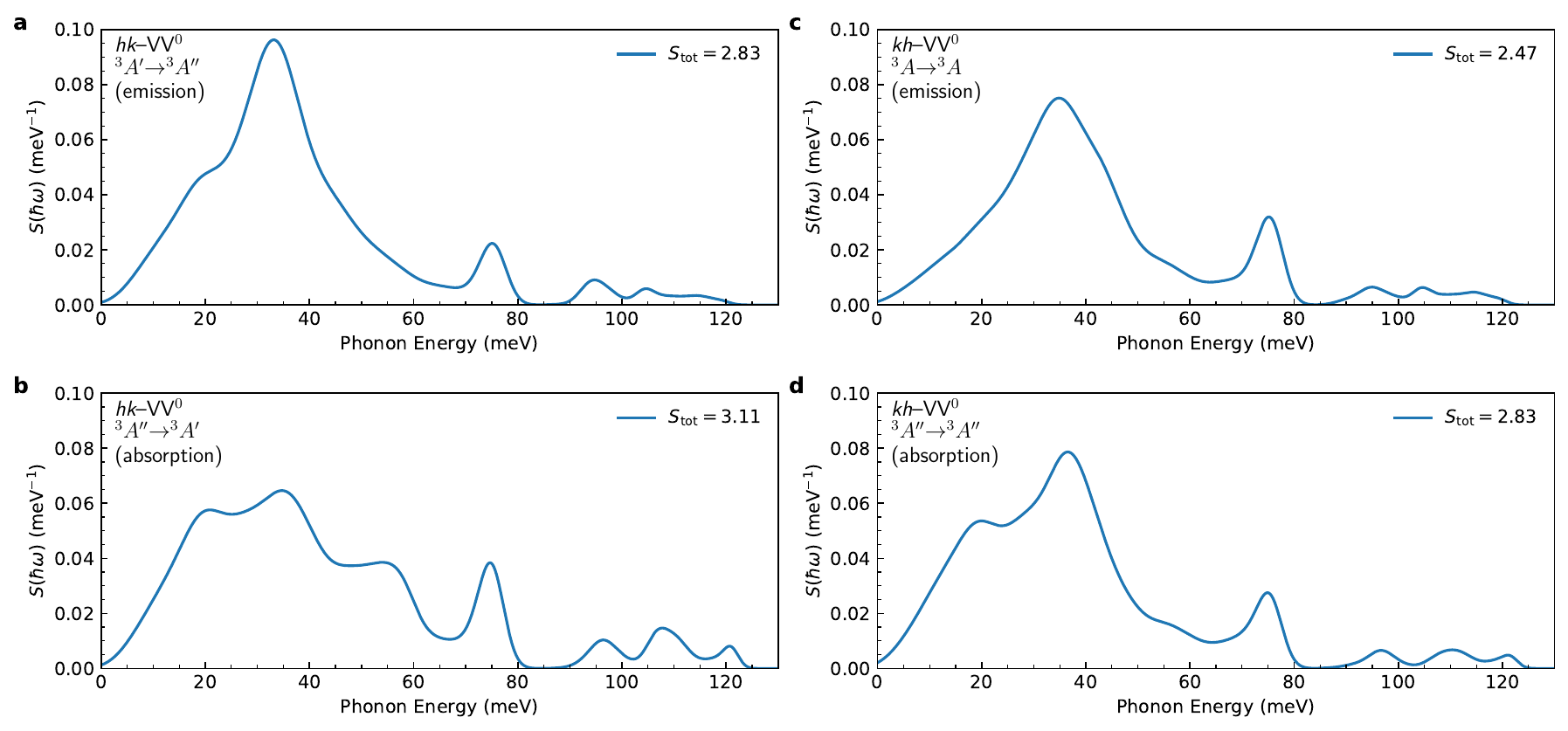}
  \caption{\textbf{Spectral densities of electron--phonon coupling for basal divacancies.} \textbf{a} Spectral function $S(\hbar\omega)$ of the ${^3A'}\to{^3A''}$ and \textbf{b} ${^3A''}\to{^3A'}$ optical transitions for $hk$ divacancy. \textbf{c} Spectral function $S(\hbar\omega)$ of the ${^3A}\to{^3A}$ and \textbf{d} ${^3A''}\to{^3A''}$ optical transitions for $kh$ divacancy. Results are for the $(23\times23\times7)$ supercell with 29\,622 atomic sites.}
  \label{fig:ep-coupling-basal}
\end{figure}

The calculated spectral densities of electron--phonon coupling for both emission and absorption processes for basal divacancies are presented in Fig.~\ref{fig:ep-coupling-basal}. Each panel includes the total HR factor~\cite{Huang_1950}, $S_\mathrm{tot}=\sum_{k}S_{k}$. Our findings indicate that the total electron--phonon coupling is noticeably stronger for $hk$ divacancies (2.83 for emission and 3.11 for absorption) compared to $kh$ divacancies (2.47 and 2.83, respectively), with the main difference lying in the spectral region around 35~\text{meV}.

By comparing the spectral functions between the emission and absorption processes, a noticeable difference can be observed, which arises from changes in the vibrational structures of the ground and excited states. For axial divacancies, the main distinction lies in the vibrational structure of JT-active modes (see Figs.~\ref{fig:ep-coupling-axial}c and \ref{fig:ep-coupling-axial}d), where a significant shift of electron--phonon coupling to higher vibrational frequencies is observed during absorption. A similar, but less pronounced, effect is observed for basal divacancies (Fig.~\ref{fig:ep-coupling-basal}). The electron--phonon coupling was calculated using the equal mode approximation, which assumes identical vibrational structures in both states. Although this assumption is not entirely accurate, for low-temperature spectra, a good approximation is to use the vibrational structure of the final state in the transition~\cite{Razinkovas_2021}. However, this change in vibrational structure indicates a quadratic electron--phonon coupling on top of the DJT effect, which we do not account for in the following analysis. This could have important implications for modeling temperature effects and temperature broadening of the ZPL in future works.

\subsection{Optical spectra of emission}

\begin{figure*}
  \includegraphics[width=\textwidth]{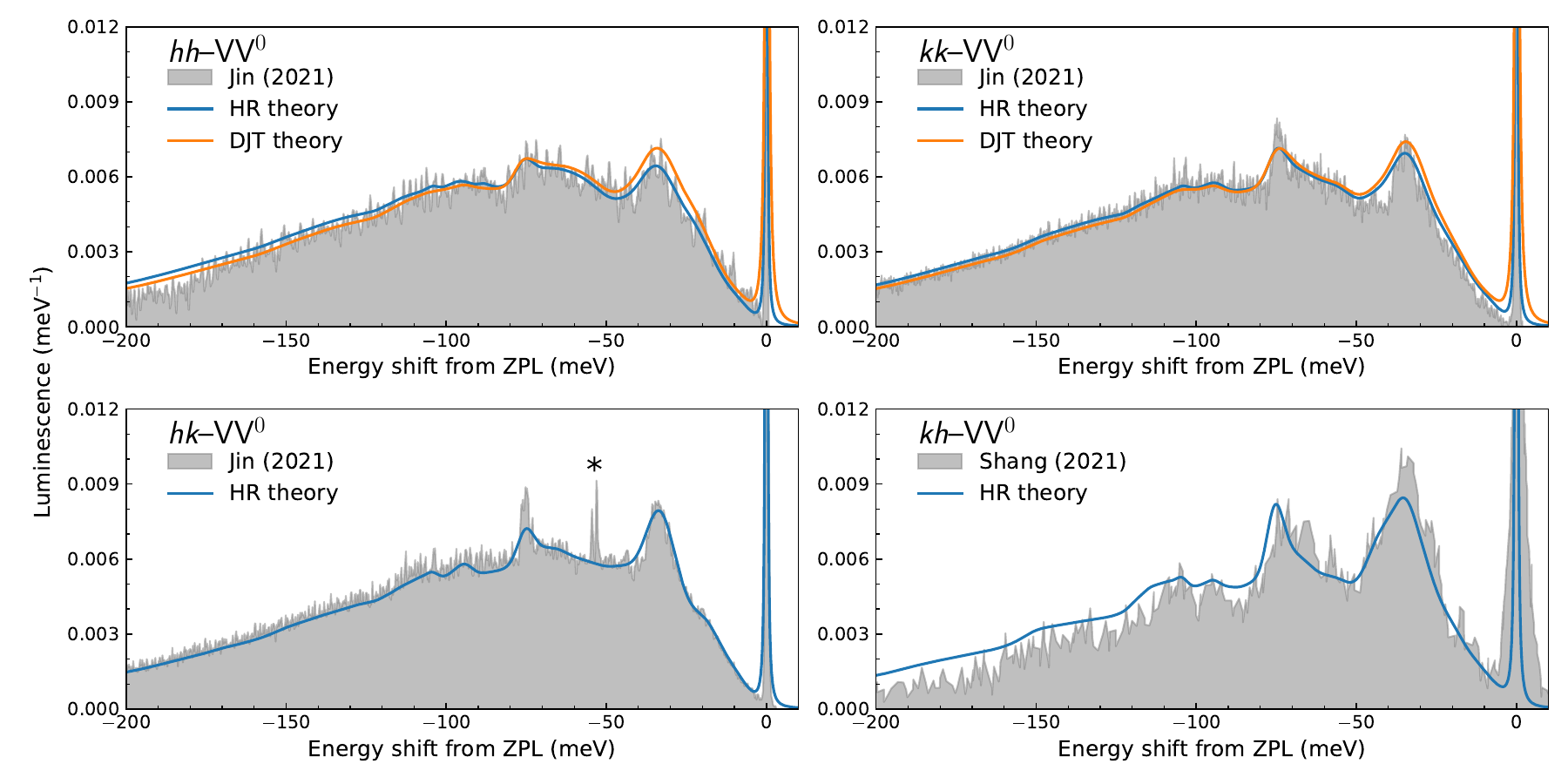}
  \caption{\textbf{Optical spectra of emission.} Blue lines are luminescence lineshapes calculated using HR theory, while the orange lines are lineshapes calculated using DJT theory. The gray area represents experimental data from Ref.~\cite{Jin_2021,Shang_2021}. The small peak marked with a star "$\ast$" in the experimental curve for \textit{hk}-VV$^{0}$ is the ZPL of \textit{hh}-VV$^{0}$ and should be disregarded in the comparison between theory and experiment. The intensities of the experimental lineshapes have been scaled to match the peaks of the computed lineshapes. Results reported here are computed at the r$^{2}$SCAN level of theory and extrapolated to the dilute limit, approximated by a $(23\times23\times7)$ supercell with 29\,622 atomic sites.}
  \label{fig:luminescence}
\end{figure*}

The ${^3E}\to{^3A_{2}}$ optical transition lineshapes calculated using HR theory and multimode DJT theory (in the case of axial divacancies) are presented in Fig.~\ref{fig:luminescence} and compared with the experimental lineshapes (in gray)~\cite{Jin_2021,Shang_2021}. The emission lineshapes calculated using both theories are in very good agreement with the experiment and capture all the spectral features. The similarity of the lineshapes obtained by both theories agrees with the findings of Ref.~\cite{Razinkovas_2021}, where it was shown that for an ${^3E}\to{^3A_{2}}$ optical transition and $K^2\approx 1$, the HR theory provides a quantitatively similar description to the rigorous DJT theory. The main difference between the emission lineshapes calculated using HR and DJT theories lies in the intensity redistribution. The DJT approach yields a higher ZPL intensity and, subsequently, a higher Debye--Waller factor (DWF), defined as the ratio of the ZPL to the entire lineshape (see Table~\ref{tab:DWF}), with a sideband that is less extended compared to that predicted by HR theory.

The $hh$, $kk$, and $hk$ divacancies exhibit similar emission lineshape profiles and DWFs. In contrast, the $kh$ divacancy shows significantly different characteristics when compared to the other divacancies. The computed mass-weighted displacements $\Delta Q$ ($\mathrm{amu}^{0.5}$ \AA) between the equilibrium structures of the ground state and the excited state are as follows: 0.823 ($hh$), 0.802 ($kk$), 0.798 ($hk$), and 0.762 ($kh$). Moreover, the total electron--phonon coupling factors are 3.21 ($hh$), 3.01 ($kk$), 2.83 ($hk$), and 2.47 ($kh$). The noticeably lower mass-weighted displacement and total electron--phonon coupling for the $kh$ divacancy lead to its more prominent DWF. These distinctions underscore the unique vibrational characteristics of the $kh$ configuration along with different polarization properties, setting it apart from the $hh$, $kk$, and $hk$ configurations.

\begin{table}
  \caption{\label{tab:DWF}Computed Debye--Waller factor (DWF) (\%) of emission computed using different levels of theory for the four different neutral divacancy configurations. The results are compared with experimental values.}
  \begin{ruledtabular}
    \begin{tabular}{lccc}
      Configuration & Calc. HR & Calc. DJT & Expt. \\
      \hline
      \textit{hh} & 5.60 & 7.04 & 3.69~\cite{Jin_2021} \\
      \textit{kk} & 6.82 & 7.03 & 6.11~\cite{Jin_2021} \\
      \textit{hk} & 7.84 &  $-$ & 7.54~\cite{Jin_2021} \\
      \textit{kh} &11.22 &  $-$ & 13~\cite{Shang_2021} \\
    \end{tabular}
  \end{ruledtabular}
\end{table}

\begin{figure*}
  \includegraphics[width=\textwidth]{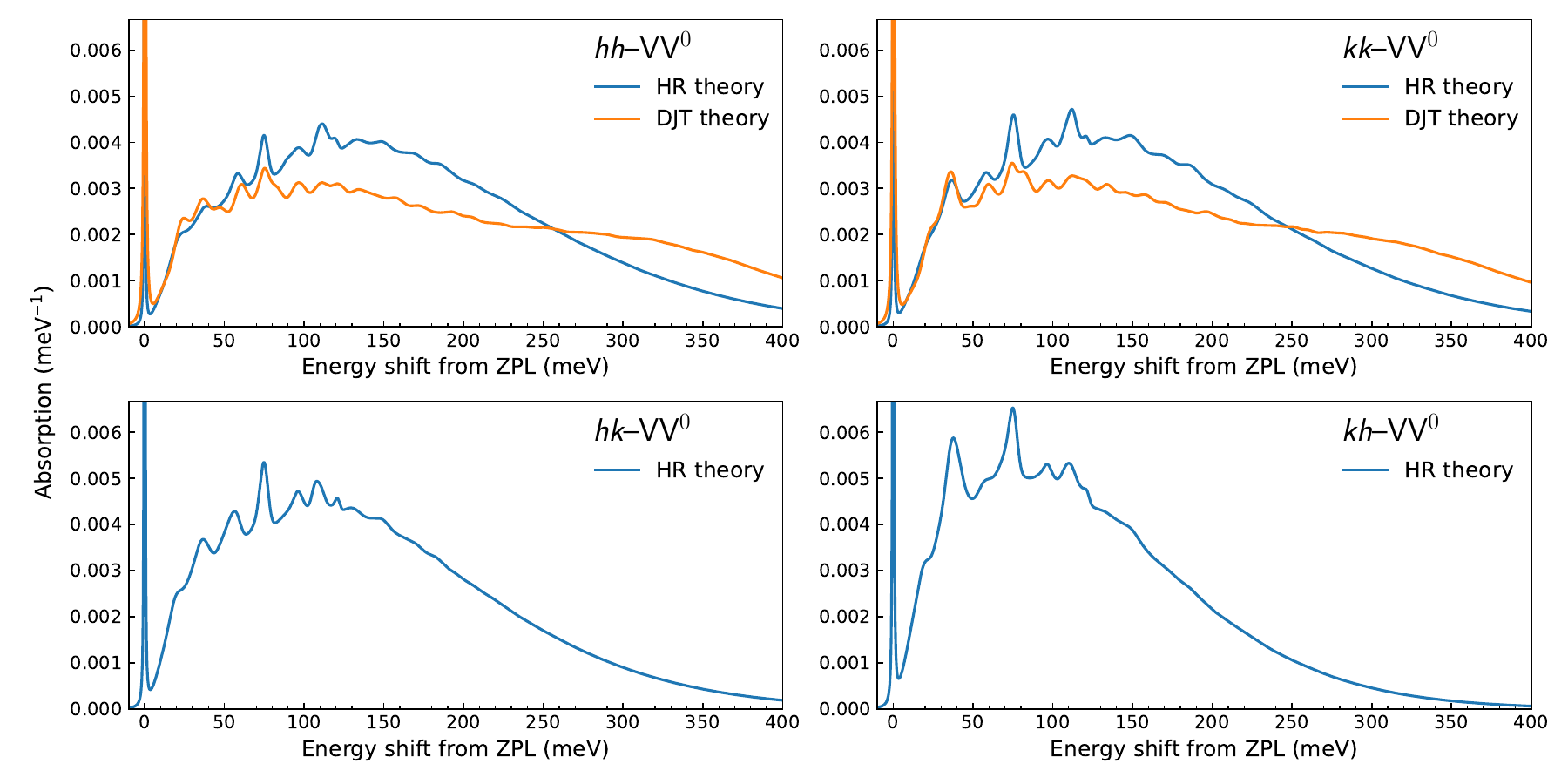}
  \caption{\textbf{Optical spectra of absorption.} Blue lines are absorption lineshapes calculated using HR theory, while the orange lines are absorption lineshapes calculated using DJT theory. Results reported here are computed at the r$^{2}$SCAN level of theory and extrapolated to the dilute limit, approximated by a $(23\times23\times7)$ supercell with 29\,622 atomic sites.}
  \label{fig:absorption}
\end{figure*}

\subsection{Optical spectra of absorption}

The absorption lineshapes for the ${^3A_{2} \to {^3E}}$ optical transition, calculated using both HR and multimode DJT theories for axial divacancies, are displayed in Fig.~\ref{fig:absorption}. Additionally, the absorption lineshapes for the ${^3A'' \to {^3A'}}$ transition in the $hk$ divacancy and the ${^3A'' \to {^3A''}}$ transition in the $kh$ divacancy, calculated with HR theory, are presented in the same figure.

Comparing the HR and DJT calculations, the DJT theory predicts a higher ZPL intensity along with more extended sidebands than HR theory, a difference that becomes particularly significant beyond 250~meV above the ZPL. Across all absorption spectra, sharp peaks near 75~meV arise from quasi-localized phonon modes. In the $kh$ divacancy, an additional prominent peak at 38~meV, resulting from delocalized bulk-like vibrations, exhibits a more pronounced intensity and a sideband that decays faster than in other divacancies. The total electron--phonon coupling factors are calculated to be 3.54 ($hh$), 3.37 ($kk$), 3.11 ($hk$), and 2.83 ($kh$). The DWFs computed using HR theory are 2.53\% ($hh$), 3.02\% ($kk$), 4.02\% ($hk$), and 5.39\% ($kh$). For the DJT model, the calculated DWFs for the $hh$ and $kk$ divacancies are slightly higher, at 3.42\% and 3.72\%, respectively.


\section{Discussion}

In summary, first-principles calculations were employed to examine the optical properties of neutral divacancies in the 4H-SiC polytype, with particular attention to the DJT effect present in the excited state of axial configurations. The approach used herein enables accurate predictions of emission and absorption spectra in the dilute limit using the force constant embedding methodology, filling gaps left by previous works that did not explicitly address the DJT effect or calculate absorption lineshapes for all configurations. Furthermore, our findings demonstrate that the r$^{2}$SCAN functional provides accurate predictions of the electronic, vibrational and vibronic properties, as reflected in the strong agreement between calculated and experimental ZPL, ZFS, DWF values and lineshapes.

During the ${^3E}\to{^3A_{2}}$ optical transition for axial divacancies, where degeneracy is present in the initial state, the lineshapes calculated using HR theory closely match those obtained with the more rigorous multimode DJT approach. In contrast, for the ${^3A_{2}}\to{^3E}$ optical transition to the degenerate excited state, multimode DJT treatment becomes essential to accurately capture the complex optical features that are characteristic of systems with strong JT interactions, such as the emission lineshape in the negatively charged nickel-vacancy center in diamond. Furthermore, the DWFs derived from multimode DJT theory predict a higher ZPL intensity with more extended sidebands compared to HR theory, particularly beyond 250~meV above the ZPL. This difference highlights the critical role of DJT effects in accurately describing optical transitions involving degenerate states.

Interestingly, our results predict spontaneous symmetry breaking in the excited state of the $kh$ divacancy due to the PJT effect. Compared to the other configurations, the $kh$-VV$^0$ was found to have the lowest total electron--phonon coupling, the highest DWF, and unique polarization characteristics, making it selectively excitable in applications involving an ensemble of divacancies.

The methodologies discussed herein not only advance understanding of the neutral divacancy's optical properties in 4H-SiC, but also illustrate a broader framework for applying multimode DJT theory to other defects with strong vibronic coupling. By addressing both emission and absorption phenomena with multimode DJT theory, the present study paves the way for exploring electron--phonon interactions and optical properties in other spin defects with pronounced JT effect, informing future designs for quantum emitters, sensors and related applications in quantum technologies.


\section{Methods}

\subsection{Electronic structure calculations}

Calculations have been performed within the spin-polarized DFT framework. Projector-augmented wave (PAW) approach was used with a plane-wave energy cutoff of 600~eV to describe the interaction of electrons with atomic cores. All calculations were performed using the Vienna \textit{Ab initio} Simulation Package (VASP) with the \textit{meta}-GGA class r$^{2}$SCAN functional~\cite{VASP,Furness_2020}. This functional was chosen due to its accuracy in describing the APES for covalent materials like diamond, SiC, and Si compared to both GGA and hybrid functionals~\cite{Maciaszek_2023}. Furthermore, when calculating optical excitation energies, r$^{2}$SCAN yields results with accuracy that is close to the much more computationally expensive hybrid or beyond-RPA methods~\cite{Maciaszek_2023,Ivanov_2023,Maciaszek_2024,Silkinis_2024}.

Divacancy defects were modeled using the supercell approach~\cite{Freysoldt_2014}. All neutral divacancy (V$_{\text{Si}}$V$_{\text{C}}^{0}$ or $\text{VV}^{0}$) centers were created by removing one Si and one C atom from a $(6\times6\times2)$ supercell with 576 atomic sites. Brillouin zone sampling was done using a single \textbf{k}-point (at $\Gamma$). The stopping criterion for the electronic self-consistent loop and ionic relaxation were set to \SI{e-8}{\electronvolt} and \SI{e-4}{\electronvolt\per\angstrom}, respectively.

\subsection{Zero-field splitting and optical excitation energy calculations}

The methodology of Ivady \textit{et al}. implemented in the VASP code was used to calculate the ZFS parameters~\cite{Ivady_2014}. The ZPL energies were calculated utilizing the $\Delta$-self consistent field or $\Delta$\text{SCF} approximation, the accuracy of which has been demonstrated in various instances when applied to defects in solids~\cite{Jones_1989,Hellman_2004,Kowalczyk_2011,Gali_2009,Davidsson_2018,Thiering_2018,Mackoit_2019,Hashemi_2021,Jin_2021}. The excited states are obtained by constraining the occupancies of the Kohn--Sham single-particle levels by promoting one electron from the highest occupied level in the spin-minority channel to the next higher energy level in the same spin channel. In the case of $hh$ and $kk$ divacancies, the excited triplet state $^{3}E$ has an electronic configuration $a_{1}e^{3}$ and is an $E\otimes(e\oplus e \cdots)$ JT system. To address computational issues stemming from the JT distortion in the excited state, caused by the degeneracy between the $a_{1}e_{x}^{1}e_{y}^{2}$ and $a_{1}e_{x}^{2}e_{y}^{1}$ configurations, the charge density was approximated by maintaining an electronic configuration of $a_{1}e_{x}^{1.5}e_{y}^{1.5}$. This practical approach helps limit the excited state density to an average between the two degenerate configurations while preserving the $C_{3v}$ symmetry.

\subsection{Vibronic states and Jahn--Teller Hamiltonian}

\def\Q{\mathbf{Q}}

In this section, we focus exclusively on the vibronic states of axial divacancies exhibiting $C_{3v}$ symmetry. For a non-degenerate electronic state, the vibronic state is expressed within the adiabatic approximation as
\begin{equation}
\ket{\Psi_{kl}} = \vwf[i][l][a_1][a_1] \vwf[i][k][e][e] \ket{A},
\label{WF1}
\end{equation}
where $\vwf[i][l][a_1][a_1]$ and $\vwf[i][k][e][e]$ represent the harmonic components of the vibrational wave function, corresponding to symmetry-preserving ($a_1$) and symmetry-breaking ($e$) vibrational modes, respectively. These components are derived from standard adiabatic phonon analysis, where $l$ and $k$ label the respective harmonic excitations. The symmetry-adapted normal coordinates $\mathbf{Q}$ collectively encapsulate the vibrational degrees of freedom. The electronic wave function, denoted by $\ket{A}$, depends solely on the electronic degrees of freedom within the static adiabatic approximation, remaining decoupled from vibrational dynamics in this formalism.

For an $E \otimes (e \oplus e \oplus \cdots)$ system, the vibronic Hamiltonian describing the linear JT interaction between an orbital doublet and $e$-symmetry vibrational modes can be expressed as $\hat{\mathcal{H}} = \hat{\mathcal{H}}_0 + \hat{\mathcal{H}}_{\mathrm{JT}}$. In the basis of electronic states $|E_{x}\rangle$ and $|E_{y}\rangle$, the zero-order Hamiltonian describes vibrational motion within a harmonic potential and is given by
\begin{align}
  \hat{\mathcal{H}}_{\mathrm{0}} = \sum_{k;\gamma\in \{x, y\}}
  \left(
    -\frac{\hbar^2}{2}\frac{\partial^2}{\partial Q^2_{k\gamma}}
    + \frac{1}{2}  \omega_k^2 Q_{k\gamma}^2
  \right)
  \begin{pmatrix}
    1 & 0 \\ 0 & 1
  \end{pmatrix}.
\label{eq:Hph}
\end{align}
The linear JT coupling term is~\cite{ham1968,Bersuker_2006}
\begin{equation}
  \hat{\mathcal{H}}_{\mathrm{JT}} = \sum_{k}
  \sqrt{2\hbar\omega_{k}^{3}}K_k
  \begin{pmatrix}
      Q_{k,y} & Q_{k,x} \\ Q_{k,x} & -Q_{k,y}
  \end{pmatrix},
\label{eq:JT}
\end{equation}
where $\omega_k$ are the angular frequencies of vibrations, derived as eigenvalues of the zero-order Hamiltonian [Eq.~\eqref{eq:Hph}], and $K_k$ are the dimensionless vibronic coupling constants~\cite{Obrien_1980}. The index $k = 1, \dots, N$ encompasses all pairs of degenerate $e$-symmetry vibrations, which transform according to the Cartesian $xy$ representation, analogous to the transformation properties of the electronic states. The solutions of the vibronic Hamiltonian are
\begin{equation}
\ket{\Phi_{m}} =
\vwf[j][m][e;x][e]\ket{E_{x}} +
\vwf[j][m][e;y][e]\ket{E_{y}},
\label{eq:vibronic_term}
\end{equation}
where the coefficients $\vwf[j][m][e;\gamma][e]$ are ionic prefactors determined by solving the JT problem, and the quantum number $m$ labels the vibronic excitations of the system.

The complete wave function, including the symmetry-preserving modes, within the degenerate electronic manifold corresponding to the JT-active state, is given as~\cite{Razinkovas_2021,Silkinis_2024}
$
\ket{\Psi_{nm}} = \vwf[j][n][a][a]\ket{\Phi_{j;m}},
\label{WF2}
$ where $n$ labels the harmonic states associated with the symmetry-preserving modes.

The solution to the DJT problem involves three essential steps:

\textbf{1. Calculation of the zero-order vibrational structure.} This step requires solving the zero-order Hamiltonian [Eq.~\eqref{eq:Hph}]. To suppress the JT interactions, the electronic configuration $a_{1}e_{x}^{1.5}e_{y}^{1.5}$ is used, following the methodology outlined in Refs.~\cite{Razinkovas_2021,Silkinis_2024}. This configuration effectively models an ensemble of two degenerate electronic states with equal probability, ensuring that the $C_{3v}$ symmetry is preserved. By maintaining this symmetry, the vibrational modes can be systematically classified according to the
irreducible representations of the $C_{3v}$ point group.

\textbf{2. Estimation of the vibronic coupling parameters.} The vibronic coupling constants, $K_k$, are extracted by analyzing the geometry change $\Delta \mathbf{Q}_{\mathrm{JT}}$ associated with the transition from the high-symmetry to the low-symmetry atomic configuration, as detailed in Supplementary Note 4. These constants are calculated by projecting $\Delta \mathbf{Q}_{\mathrm{JT}}$ onto a pair $k$ of $e$-symmetry normal modes, according to the expression~\cite{Razinkovas_2021}
\begin{equation}
  K^2_k = \frac{\omega_k \Delta Q_{k}^2}{2\hbar},
\end{equation}
where $\Delta Q_k^2 = \Delta Q_{kx}^2 + \Delta Q_{ky}^2$ represents the squared projection of the $k$-doublet normal coordinates along the relaxation path $\Delta \mathbf{Q}_{\mathrm{JT}}$. For embedded supercells, recovering $\Delta \mathbf{Q}_{\mathrm{JT}}$ in the dilute limit is achieved through force-based methods, as described in Refs.~\cite{Razinkovas_2021,Silkinis_2024}.

\textbf{3. Diagonalizing the vibronic Hamiltonian.} The diagonalization of the Hamiltonian $\mathcal{H} = \mathcal{H}_0 + \mathcal{H}_{\text{JT}}$, which includes many vibrational modes, poses significant computational challenges due to the large matrix dimensions. To manage this complexity, we adopt an approach based on a reduced set of effective modes \cite{Razinkovas_2021,Silkinis_2024}. In this method, the spectral density of the JT coupling is defined as $K^2(\hbar\omega) = \sum_k K_k^2 \delta(\hbar\omega - \hbar\omega_k)$. This density is then approximated by a density stemming from a smaller number of effective modes $ K^2_{\mathrm{eff}}(\hbar\omega) = \sum_{n=1}^{N_{\mathrm{eff}}} \bar{K}^2_{n} g_{\sigma}(\hbar\bar{\omega}_{n} - \hbar\omega)$, where $g_{\sigma}$ is a Gaussian function with width $\sigma$. The approximation uses $N_{\mathrm{eff}}$ effective vibrations, each characterized by a frequency $\bar{\omega}_n$ and an associated vibronic coupling strength $\bar{K}^2_{n}$. The parameters $\bar{K}^2_n$, $\bar{\omega}_{n}$, and $\sigma$ are optimized to ensure that $K^2_{\mathrm{eff}}(\hbar\omega)$ accurately reproduces $K^2(\hbar\omega)$. This reduction significantly lowers the computational burden, as the number of effective modes is much smaller than the total number of modes, $N_{\mathrm{eff}} \ll N$.

To validate the reliability of the spectral functions derived from this approximation, we systematically monitor their convergence with respect to the number of effective modes. Details on the convergence analysis are provided in Supplementary Note 6.

\subsection{Phonon calculations}

Phonon modes of divacancies were calculated using the finite displacement method with configurations generated using the \texttt{PHONOPY} software package~\cite{Togo_2015}. Displacements of $\pm 0.01\text{ \AA}$ from equilibrium geometries were used. Extrapolation of phonon modes to the dilute limit was done with a $(23\times23\times7)$ supercell with 29\,622 atomic sites using the force constant matrix embedding methodology presented in Refs.~\cite{Alkauskas_2014,Razinkovas_2021}. To achieve a smooth representation of the electron--phonon spectral function, delta functions were approximated with Gaussian functions of variable widths ($\sigma$), decreasing linearly from 3.5~meV at zero frequency to 1.5~meV at the highest phonon energy.

\subsection{Calculation of optical lineshapes}

In the zero-temperature limit ($T=0$~K) of the Franck--Condon approximation, the
normalized lineshape $L(\hbar\omega)$ for absorption and luminescence is given
by~\cite{Razinkovas_2021}
\begin{eqnarray}
  \label{eq:P}
  L(\hbar\omega) = C\omega^{\kappa}A(\hbar\omega),
\end{eqnarray}
where $C$ is a normalization constant, $A(\hbar\omega)$ is the spectral function
describing the phonon sideband, and $\kappa$ takes the value of 3 for emission
and 1 for absorption. For transitions involving a single degenerate orbital
doublet, the spectral function is expressed as a
convolution~\cite{Razinkovas_2021}:
\begin{equation}
  A(\hbar\omega) =
  \int
  A_{a_{1}}(\hbar\omega - \hbar\omega')
  A_{e}(\hbar\omega')
  \mathrm{d}(\hbar\omega'),
  \label{conv}
\end{equation}
where $A_{a_1}$ and $A_e$ are the spectral functions corresponding to
symmetry-preserving ($a_1$) and symmetry-breaking ($e$) vibrational modes,
respectively. The $a_1$ spectral function is given by
\begin{align}
  A_{a_1}(\hbar\omega) &=
  \sum_n
  \left|
  \braket{\chi_{i 0}^{a_1}}{\chi_{f n}^{a_1}}
  \right|^2
  \delta\left(
  E_{\mathrm{ZPL}} \mp (\varepsilon_{f n}^{a_1} - \varepsilon_{f 0}^{a_1})
  - \hbar\omega
  \right),
  \label{eq:spectral_fa1}
\end{align}
where the minus sign applies to emission, the plus sign to absorption,
$i$ and $f$ refer to the initial and final electronic states, and
$\varepsilon_{f n}^{a_1}$ is the energy eigenvalue of the $n$-th vibrational
level in state $f$.

The spectral function for JT active $e$ vibrational modes is
\begin{align}
  A_e(\hbar\omega) =
  & \sum_m
    \left[
    \left|\braket{\vwf*[e][i0][e;x][e]}{\vwf*[g][fm][e][e]}\right|^2 +
    \left|\braket{\vwf*[e][i0][e;y][e]}{\vwf*[g][fm][e][e]}\right|^2
    \right]
    \delta(\varepsilon^{e}_{i0} - \varepsilon^e_{im} - \hbar\omega)
\label{e-lum}
\end{align}
for emission, whereby the initial state is degenerate, and
\begin{align}
  A_e(\hbar\omega) &=
  \sum_m
  \bigl[
  \left|\braket{\chi_{i 0}^{e}}{\chi_{f m}^{e;x}}\right|^2 +
  \left|\braket{\chi_{i 0}^{e}}{\chi_{f m}^{e;x}}\right|^2
  \bigr]
  \delta(\varepsilon_{f m}^{e} - \varepsilon_{f 0}^{e} - \hbar\omega)
  \label{e-abs}
\end{align}
for absorption, whereby the degeneracy is in the final state. Here, the $x$ and
$y$ superscripts refer to ionic prefactors in Eq.~\eqref{eq:vibronic_term}, and
$\varepsilon_{f m}^{e}$ represents the energy of the $m$-th vibronic quantum
level in the electronic state $f$.

The spectral function for symmetry-preserving $a_1$ modes
[Eq.~\eqref{eq:spectral_fa1}] was computed using the HR theory and
the method of generating functions. This approach involves calculating partial
HR factors $S_k$ and the spectral density of electron–phonon coupling, which is defined as
$S(\hbar\omega) = \sum_k S_k \delta(\hbar\omega - \hbar\omega_k)$. The spectral
function $A_{a_1}$ is then derived from this spectral density using a time-domain
generating function formalism. The methodology for calculating $S_k$ and
$A_{a_1}$ is well established and described in detail in
Refs.~\cite{Alkauskas_2014,Razinkovas_2021,Silkinis_2024b}.

For the degenerate $e$-symmetry modes [Eqs.~\eqref{e-lum} and~\eqref{e-abs}],
the spectral function was obtained by explicitly solving the vibronic problem in
the form of Eq.~\eqref{eq:vibronic_term} and evaluating the overlaps between the
vibrational wave functions described by Eqs.~\eqref{WF1}
and~\eqref{eq:vibronic_term}. The delta functions in Eqs.~\eqref{e-lum}
and~\eqref{e-abs} were replaced with Gaussian functions of width 3.5~meV to
account for finite broadening. For benchmarking purposes, we also computed the
spectral function for emission pertaining to JT active $e$ modes
using the HR theory with Eq.~\eqref{eq:spectral_fa1}, instead of
Eq.~\eqref{e-lum}. In this approach, the contribution of $e$ modes is
approximated by treating the minimum of APES for $e$ degrees of freedom in the initial state as the minimum of harmonic potential. This approximation enables the estimation of HR
parameters. Although this method is not rigorous, it offers a reasonable
approximation within zero-temperature theory for an initial state that is JT
active, provided the total vibronic coupling strength, $K_{\mathrm{tot}}^{2}$,
lies within the range of about 0.5–1~\cite{Razinkovas_2021}.

\section{Acknowledgements}
Financial support was kindly provided by the Research Council of Norway and the University of Oslo through the frontier research project QuTe (no. 325573, FriPro ToppForsk-program).
The work of MEB was supported by an ETH Zürich Postdoctoral Fellowship. Computational resources were provided by supercomputer GALAX of the Center for Physical Sciences and Technology, Lithuania, and the High Performance Computing Center “HPC Saulėtekis” in the Faculty of Physics, Vilnius University.

\section{Author contributions}

LR, LV, and MEB designed the research. VŽ, RS, and LR wrote the codes and performed the electron--phonon coupling calculations with supervision by MEB and LR. VŽ performed the ZPL, ZFS, APES, and DWF calculations with supervision by MEB and LR. VŽ and LR analyzed the data. VŽ, MEB, and LR wrote the manuscript with support and editing from all authors.

\section{Competing interests}
The authors declare no competing interests.

\bibliography{main}

\begin{thebibliography}{10}
\expandafter\ifx\csname url\endcsname\relax
  \def\url#1{\texttt{#1}}\fi
\expandafter\ifx\csname urlprefix\endcsname\relax\def\urlprefix{URL }\fi
\providecommand{\bibinfo}[2]{#2}
\providecommand{\eprint}[2][]{\url{#2}}

\bibitem{She_2017}
\bibinfo{author}{She, X.}, \bibinfo{author}{Huang, A.~Q.},
  \bibinfo{author}{Lucia, O.} \& \bibinfo{author}{Ozpineci, B.}
\newblock \bibinfo{title}{Review of silicon carbide power devices and their
  applications}.
\newblock \emph{\bibinfo{journal}{IEEE Transactions on Industrial Electronics}}
  \textbf{\bibinfo{volume}{64}}, \bibinfo{pages}{8193--8205}
  (\bibinfo{year}{2017}).

\bibitem{Kimoto_2014}
\bibinfo{author}{Kimoto, T.} \& \bibinfo{author}{Cooper, J.~A.}
\newblock \emph{\bibinfo{title}{Fundamentals of silicon carbide technology:
  growth, characterization, devices and applications}}
  (\bibinfo{publisher}{John Wiley \& Sons}, \bibinfo{year}{2014}).

\bibitem{Klein_2006}
\bibinfo{author}{Klein, P.} \emph{et~al.}
\newblock \bibinfo{title}{Lifetime-limiting defects in n- 4{H}-{SiC}
  epilayers}.
\newblock \emph{\bibinfo{journal}{Applied Physics Letters}}
  \textbf{\bibinfo{volume}{88}}, \bibinfo{pages}{052110}
  (\bibinfo{year}{2006}).

\bibitem{Ghezellou_2023}
\bibinfo{author}{Ghezellou, M.} \emph{et~al.}
\newblock \bibinfo{title}{The role of boron related defects in limiting charge
  carrier lifetime in 4{H}–{SiC} epitaxial layers}.
\newblock \emph{\bibinfo{journal}{APL Materials}}
  \textbf{\bibinfo{volume}{11}}, \bibinfo{pages}{031107}
  (\bibinfo{year}{2023}).

\bibitem{Aharonovich_2016}
\bibinfo{author}{Aharonovich, I.}, \bibinfo{author}{Englund, D.} \&
  \bibinfo{author}{Toth, M.}
\newblock \bibinfo{title}{Solid-state single-photon emitters}.
\newblock \emph{\bibinfo{journal}{Nature Photonics}}
  \textbf{\bibinfo{volume}{10}}, \bibinfo{pages}{631--641}
  (\bibinfo{year}{2016}).

\bibitem{Castelletto_2020}
\bibinfo{author}{Castelletto, S.} \& \bibinfo{author}{Boretti, A.}
\newblock \bibinfo{title}{Silicon carbide color centers for quantum
  applications}.
\newblock \emph{\bibinfo{journal}{Journal of Physics: Photonics}}
  \textbf{\bibinfo{volume}{2}}, \bibinfo{pages}{022001} (\bibinfo{year}{2020}).

\bibitem{Weber_2010}
\bibinfo{author}{Weber, J.~R.} \emph{et~al.}
\newblock \bibinfo{title}{Quantum computing with defects}.
\newblock \emph{\bibinfo{journal}{Proceedings of the National Academy of
  Sciences}} \textbf{\bibinfo{volume}{107}}, \bibinfo{pages}{8513--8518}
  (\bibinfo{year}{2010}).

\bibitem{Awschalom_2018}
\bibinfo{author}{Awschalom, D.~D.}, \bibinfo{author}{Hanson, R.},
  \bibinfo{author}{Wrachtrup, J.} \& \bibinfo{author}{Zhou, B.~B.}
\newblock \bibinfo{title}{Quantum technologies with optically interfaced
  solid-state spins}.
\newblock \emph{\bibinfo{journal}{Nature Photonics}}
  \textbf{\bibinfo{volume}{12}}, \bibinfo{pages}{516--527}
  (\bibinfo{year}{2018}).

\bibitem{Zhang_2020}
\bibinfo{author}{Zhang, G.}, \bibinfo{author}{Cheng, Y.},
  \bibinfo{author}{Chou, J.-P.} \& \bibinfo{author}{Gali, A.}
\newblock \bibinfo{title}{Material platforms for defect qubits and
  single-photon emitters}.
\newblock \emph{\bibinfo{journal}{Applied Physics Reviews}}
  \textbf{\bibinfo{volume}{7}}, \bibinfo{pages}{031308} (\bibinfo{year}{2020}).

\bibitem{Son_2020}
\bibinfo{author}{Son, N.~T.} \emph{et~al.}
\newblock \bibinfo{title}{Developing silicon carbide for quantum spintronics}.
\newblock \emph{\bibinfo{journal}{Applied Physics Letters}}
  \textbf{\bibinfo{volume}{116}}, \bibinfo{pages}{190501}
  (\bibinfo{year}{2020}).

\bibitem{Bathen_QuTe_2021}
\bibinfo{author}{Bathen, M.~E.} \& \bibinfo{author}{Vines, L.}
\newblock \bibinfo{title}{Manipulating single-photon emission from point
  defects in diamond and silicon carbide}.
\newblock \emph{\bibinfo{journal}{Advanced Quantum Technologies}}
  \textbf{\bibinfo{volume}{4}}, \bibinfo{pages}{2100003}
  (\bibinfo{year}{2021}).

\bibitem{Widmann_2014}
\bibinfo{author}{Widmann, M.} \emph{et~al.}
\newblock \bibinfo{title}{Coherent control of single spins in silicon carbide
  at room temperature}.
\newblock \emph{\bibinfo{journal}{Nature Materials}}
  \textbf{\bibinfo{volume}{14}}, \bibinfo{pages}{164--168}
  (\bibinfo{year}{2015}).

\bibitem{Castelletto_2014}
\bibinfo{author}{Castelletto, S.} \emph{et~al.}
\newblock \bibinfo{title}{A silicon carbide room-temperature single-photon
  source}.
\newblock \emph{\bibinfo{journal}{Nature Materials}}
  \textbf{\bibinfo{volume}{13}}, \bibinfo{pages}{151--156}
  (\bibinfo{year}{2014}).

\bibitem{Christle_2015}
\bibinfo{author}{Christle, D.~J.} \emph{et~al.}
\newblock \bibinfo{title}{Isolated electron spins in silicon carbide with
  millisecond coherence times}.
\newblock \emph{\bibinfo{journal}{Nature Materials}}
  \textbf{\bibinfo{volume}{14}}, \bibinfo{pages}{160--163}
  (\bibinfo{year}{2015}).

\bibitem{Christle_2017}
\bibinfo{author}{Christle, D.~J.} \emph{et~al.}
\newblock \bibinfo{title}{Isolated spin qubits in {SiC} with a high-fidelity
  infrared spin-to-photon interface}.
\newblock \emph{\bibinfo{journal}{Physical Review X}}
  \textbf{\bibinfo{volume}{7}}, \bibinfo{pages}{021046} (\bibinfo{year}{2017}).

\bibitem{Anderson_2022}
\bibinfo{author}{Anderson, C.~P.} \emph{et~al.}
\newblock \bibinfo{title}{Five-second coherence of a single spin with
  single-shot readout in silicon carbide}.
\newblock \emph{\bibinfo{journal}{Science advances}}
  \textbf{\bibinfo{volume}{8}}, \bibinfo{pages}{eabm5912}
  (\bibinfo{year}{2022}).

\bibitem{Alkauskas_2014}
\bibinfo{author}{Alkauskas, A.}, \bibinfo{author}{Buckley, B.~B.},
  \bibinfo{author}{Awschalom, D.~D.} \& \bibinfo{author}{Van~de Walle, C.~G.}
\newblock \bibinfo{title}{First-principles theory of the luminescence lineshape
  for the triplet transition in diamond {NV} centres}.
\newblock \emph{\bibinfo{journal}{New Journal of Physics}}
  \textbf{\bibinfo{volume}{16}}, \bibinfo{pages}{073026}
  (\bibinfo{year}{2014}).

\bibitem{Razinkovas_2021}
\bibinfo{author}{Razinkovas, L.}, \bibinfo{author}{Doherty, M.~W.},
  \bibinfo{author}{Manson, N.~B.}, \bibinfo{author}{Van~de Walle, C.~G.} \&
  \bibinfo{author}{Alkauskas, A.}
\newblock \bibinfo{title}{Vibrational and vibronic structure of isolated point
  defects: The nitrogen-vacancy center in diamond}.
\newblock \emph{\bibinfo{journal}{Physical Review B}}
  \textbf{\bibinfo{volume}{104}}, \bibinfo{pages}{045303}
  (\bibinfo{year}{2021}).

\bibitem{Abtew_2011}
\bibinfo{author}{Abtew, T.~A.} \emph{et~al.}
\newblock \bibinfo{title}{Dynamic {J}ahn-{T}eller effect in the {NV}$^-$ center
  in diamond}.
\newblock \emph{\bibinfo{journal}{Physical Review Letters}}
  \textbf{\bibinfo{volume}{107}}, \bibinfo{pages}{146403}
  (\bibinfo{year}{2011}).

\bibitem{Thiering_2017}
\bibinfo{author}{Thiering, G.} \& \bibinfo{author}{Gali, A.}
\newblock \bibinfo{title}{Ab initio calculation of spin-orbit coupling for an
  {NV} center in diamond exhibiting dynamic {J}ahn-{T}eller effect}.
\newblock \emph{\bibinfo{journal}{Physical Review B}}
  \textbf{\bibinfo{volume}{96}}, \bibinfo{pages}{081115(R)}
  (\bibinfo{year}{2017}).

\bibitem{Zhang_2018}
\bibinfo{author}{Zhang, J.}, \bibinfo{author}{Wang, C.-Z.},
  \bibinfo{author}{Zhu, Z.}, \bibinfo{author}{Liu, Q.~H.} \&
  \bibinfo{author}{Ho, K.-M.}
\newblock \bibinfo{title}{Multimode {Jahn}-{Teller} effect in bulk systems: A
  case of the {NV}$^0$ center in diamond}.
\newblock \emph{\bibinfo{journal}{Physical Review B}}
  \textbf{\bibinfo{volume}{97}}, \bibinfo{pages}{165204}
  (\bibinfo{year}{2018}).

\bibitem{Thiering_2018}
\bibinfo{author}{Thiering, G.} \& \bibinfo{author}{Gali, A.}
\newblock \bibinfo{title}{Ab initio magneto-optical spectrum of group-{IV}
  vacancy color centers in diamond}.
\newblock \emph{\bibinfo{journal}{Physical Review X}}
  \textbf{\bibinfo{volume}{8}}, \bibinfo{pages}{021063} (\bibinfo{year}{2018}).

\bibitem{Thiering_2019}
\bibinfo{author}{Thiering, G.} \& \bibinfo{author}{Gali, A.}
\newblock \bibinfo{title}{The $(e_g\otimes e_u)\otimes e_g$ product
  {Jahn}--{Teller} effect in the neutral group-{IV} vacancy quantum bits in
  diamond}.
\newblock \emph{\bibinfo{journal}{npj Computational Materials}}
  \textbf{\bibinfo{volume}{5}}, \bibinfo{pages}{18} (\bibinfo{year}{2019}).

\bibitem{Thiering_2021}
\bibinfo{author}{Thiering, G.} \& \bibinfo{author}{Gali, A.}
\newblock \bibinfo{title}{Magneto-optical spectra of the split nickel-vacancy
  defect in diamond}.
\newblock \emph{\bibinfo{journal}{Physical Review Research}}
  \textbf{\bibinfo{volume}{3}}, \bibinfo{pages}{043052} (\bibinfo{year}{2021}).

\bibitem{Carbery_2024}
\bibinfo{author}{Carbery, W.~P.}, \bibinfo{author}{Farfan, C.~A.},
  \bibinfo{author}{Ulbricht, R.} \& \bibinfo{author}{Turner, D.~B.}
\newblock \bibinfo{title}{The phonon-modulated {Jahn}--{Teller} distortion of
  the nitrogen vacancy center in diamond}.
\newblock \emph{\bibinfo{journal}{Nature Communications}}
  \textbf{\bibinfo{volume}{15}}, \bibinfo{pages}{8646} (\bibinfo{year}{2024}).

\bibitem{Ham_1965}
\bibinfo{author}{Ham, F.~S.}
\newblock \bibinfo{title}{Dynamical {Jahn}--{Teller} effect in paramagnetic
  resonance spectra: orbital reduction factors and partial quenching of
  spin-orbit interaction}.
\newblock \emph{\bibinfo{journal}{Physical Review}}
  \textbf{\bibinfo{volume}{138}}, \bibinfo{pages}{A1727}
  (\bibinfo{year}{1965}).

\bibitem{Longuet_1958}
\bibinfo{author}{Longuet-Higgins, H.~C.}, \bibinfo{author}{{\"O}pik, U.},
  \bibinfo{author}{Pryce, M. H.~L.} \& \bibinfo{author}{Sack, R.}
\newblock \bibinfo{title}{Studies of the {J}ahn-{T}eller effect. {II}. the
  dynamical problem}.
\newblock \emph{\bibinfo{journal}{Proceedings of the Royal Society of London.
  Series A. Mathematical and Physical Sciences}}
  \textbf{\bibinfo{volume}{244}}, \bibinfo{pages}{1--16}
  (\bibinfo{year}{1958}).

\bibitem{Bersuker_2006}
\bibinfo{author}{Bersuker, I.}
\newblock \emph{\bibinfo{title}{The Jahn-Teller Effect}}
  (\bibinfo{publisher}{Cambridge University Press}, \bibinfo{year}{2006}).

\bibitem{Silkinis_2024}
\bibinfo{author}{Silkinis, R.} \emph{et~al.}
\newblock \bibinfo{title}{Optical lineshapes for orbital singlet to doublet
  transitions in a dynamical {J}ahn-{T}eller system: the {NiV}$^{-}$ center in
  diamond}.
\newblock \emph{\bibinfo{journal}{Physical Review B}}
  \textbf{\bibinfo{volume}{110}}, \bibinfo{pages}{075303}
  (\bibinfo{year}{2024}).

\bibitem{Furness_2020}
\bibinfo{author}{Furness, J.~W.}, \bibinfo{author}{Kaplan, A.~D.},
  \bibinfo{author}{Ning, J.}, \bibinfo{author}{Perdew, J.~P.} \&
  \bibinfo{author}{Sun, J.}
\newblock \bibinfo{title}{Accurate and numerically efficient {r2SCAN}
  meta-generalized gradient approximation}.
\newblock \emph{\bibinfo{journal}{The journal of physical chemistry letters}}
  \textbf{\bibinfo{volume}{11}}, \bibinfo{pages}{8208--8215}
  (\bibinfo{year}{2020}).

\bibitem{Jin_2021}
\bibinfo{author}{Jin, Y.} \emph{et~al.}
\newblock \bibinfo{title}{Photoluminescence spectra of point defects in
  semiconductors: Validation of first-principles calculations}.
\newblock \emph{\bibinfo{journal}{Physical Review Materials}}
  \textbf{\bibinfo{volume}{5}}, \bibinfo{pages}{084603} (\bibinfo{year}{2021}).

\bibitem{Davidsson_2020}
\bibinfo{author}{Davidsson, J.}
\newblock \bibinfo{title}{Theoretical polarization of zero phonon lines in
  point defects}.
\newblock \emph{\bibinfo{journal}{Journal of Physics: Condensed Matter}}
  \textbf{\bibinfo{volume}{32}}, \bibinfo{pages}{385502}
  (\bibinfo{year}{2020}).

\bibitem{Stenlund_2023}
\bibinfo{author}{Stenlund, W.}, \bibinfo{author}{Davidsson, J.},
  \bibinfo{author}{Ivády, V.}, \bibinfo{author}{Armiento, R.} \&
  \bibinfo{author}{Abrikosov, I.~A.}
\newblock \bibinfo{title}{{ADAQ}-{SYM}: Automated symmetry analysis of defect
  orbitals} (\bibinfo{year}{2023}).
\newblock \eprint{2307.04381}.

\bibitem{Shafizadeh_2024}
\bibinfo{author}{Shafizadeh, D.} \emph{et~al.}
\newblock \bibinfo{title}{Selection rules in the excitation of the divacancy
  and the nitrogen-vacancy pair in 4{H}-and 6{H}-{SiC}}.
\newblock \emph{\bibinfo{journal}{Physical Review B}}
  \textbf{\bibinfo{volume}{109}}, \bibinfo{pages}{235203}
  (\bibinfo{year}{2024}).

\bibitem{Falk_2013}
\bibinfo{author}{Falk, A.~L.} \emph{et~al.}
\newblock \bibinfo{title}{Polytype control of spin qubits in silicon carbide}.
\newblock \emph{\bibinfo{journal}{Nature communications}}
  \textbf{\bibinfo{volume}{4}}, \bibinfo{pages}{1819} (\bibinfo{year}{2013}).

\bibitem{Sun_2015}
\bibinfo{author}{Sun, J.}, \bibinfo{author}{Ruzsinszky, A.} \&
  \bibinfo{author}{Perdew, J.~P.}
\newblock \bibinfo{title}{Strongly constrained and appropriately normed
  semilocal density functional}.
\newblock \emph{\bibinfo{journal}{Physical Review Letters}}
  \textbf{\bibinfo{volume}{115}}, \bibinfo{pages}{036402}
  (\bibinfo{year}{2015}).

\bibitem{Fu_2009}
\bibinfo{author}{Fu, K.-M.} \emph{et~al.}
\newblock \bibinfo{title}{Observation of the dynamic {J}ahn-teller {E}ffect in
  the excited states of nitrogen-vacancy centers in diamond}.
\newblock \emph{\bibinfo{journal}{Physical Review Letters}}
  \textbf{\bibinfo{volume}{103}}, \bibinfo{pages}{256404}
  (\bibinfo{year}{2009}).

\bibitem{Bian_2024}
\bibinfo{author}{Bian, G.}, \bibinfo{author}{Thiering, G.} \&
  \bibinfo{author}{Gali, A.}
\newblock \bibinfo{title}{Theory of optical spinpolarization of axial divacancy
  and nitrogen-vacancy defects in 4{H}-{SiC}} (\bibinfo{year}{2024}).
\newblock \eprint{2409.10233}.

\bibitem{Huang_1950}
\bibinfo{author}{Huang, K.} \& \bibinfo{author}{Rhys, A.}
\newblock \bibinfo{title}{Theory of light absorption and non-radiative
  transitions in {F}-centres}.
\newblock \emph{\bibinfo{journal}{Proceedings of the Royal Society of London.
  Series A. Mathematical and Physical Sciences}}
  \textbf{\bibinfo{volume}{204}}, \bibinfo{pages}{406--423}
  (\bibinfo{year}{1950}).

\bibitem{Obrien_1980}
\bibinfo{author}{{O'Brien}, C.~M., Mary} \& \bibinfo{author}{Evangelou, S.~N.}
\newblock \bibinfo{title}{The calculation of absorption band shapes in dynamic
  {Jahn}--{Teller} systems by the use of the {Lanczos} algorithm}.
\newblock \emph{\bibinfo{journal}{J. Phys. C: Solid State Phys.}}
  \textbf{\bibinfo{volume}{13}}, \bibinfo{pages}{611--623}
  (\bibinfo{year}{1980}).

\bibitem{Shang_2021}
\bibinfo{author}{Shang, Z.} \emph{et~al.}
\newblock \bibinfo{title}{Microwave-assisted spectroscopy of vacancy-related
  spin centers in hexagonal {SiC}}.
\newblock \emph{\bibinfo{journal}{Physical Review Applied}}
  \textbf{\bibinfo{volume}{15}}, \bibinfo{pages}{034059}
  (\bibinfo{year}{2021}).

\bibitem{VASP}
\bibinfo{author}{Kresse, G.} \& \bibinfo{author}{Furthm{\"u}ller, J.}
\newblock \bibinfo{title}{Efficient iterative schemes for ab initio
  total-energy calculations using a plane-wave basis set}.
\newblock \emph{\bibinfo{journal}{Physical Review B}}
  \textbf{\bibinfo{volume}{54}}, \bibinfo{pages}{11169} (\bibinfo{year}{1996}).

\bibitem{Maciaszek_2023}
\bibinfo{author}{Maciaszek, M.}, \bibinfo{author}{{\v{Z}}alandauskas, V.},
  \bibinfo{author}{Silkinis, R.}, \bibinfo{author}{Alkauskas, A.} \&
  \bibinfo{author}{Razinkovas, L.}
\newblock \bibinfo{title}{The application of the {SCAN} density functional to
  color centers in diamond}.
\newblock \emph{\bibinfo{journal}{The Journal of Chemical Physics}}
  \textbf{\bibinfo{volume}{159}}, \bibinfo{pages}{084708}
  (\bibinfo{year}{2023}).

\bibitem{Ivanov_2023}
\bibinfo{author}{Ivanov, A.~V.}, \bibinfo{author}{Schmerwitz, Y. L.~A.},
  \bibinfo{author}{Levi, G.} \& \bibinfo{author}{Jónsson, H.}
\newblock \bibinfo{title}{Electronic excitations of the charged
  nitrogen-vacancy center in diamond obtained using time-independent
  variational density functional calculations}.
\newblock \emph{\bibinfo{journal}{SciPost Phys.}}
  \textbf{\bibinfo{volume}{15}}, \bibinfo{pages}{009} (\bibinfo{year}{2023}).

\bibitem{Maciaszek_2024}
\bibinfo{author}{Maciaszek, M.} \& \bibinfo{author}{Razinkovas, L.}
\newblock \bibinfo{title}{Blue quantum emitter in hexagonal boron nitride and a
  carbon chain tetramer: a first-principles study}.
\newblock \emph{\bibinfo{journal}{ACS Applied Nano Materials}}
  \textbf{\bibinfo{volume}{7}}, \bibinfo{pages}{18979--18985}
  (\bibinfo{year}{2024}).

\bibitem{Freysoldt_2014}
\bibinfo{author}{Freysoldt, C.} \emph{et~al.}
\newblock \bibinfo{title}{First-principles calculations for point defects in
  solids}.
\newblock \emph{\bibinfo{journal}{Reviews of modern physics}}
  \textbf{\bibinfo{volume}{86}}, \bibinfo{pages}{253} (\bibinfo{year}{2014}).

\bibitem{Ivady_2014}
\bibinfo{author}{Iv{\'a}dy, V.}, \bibinfo{author}{Simon, T.},
  \bibinfo{author}{Maze, J.~R.}, \bibinfo{author}{Abrikosov, I.} \&
  \bibinfo{author}{Gali, A.}
\newblock \bibinfo{title}{Pressure and temperature dependence of the zero-field
  splitting in the ground state of {NV} centers in diamond: A first-principles
  study}.
\newblock \emph{\bibinfo{journal}{Physical Review B}}
  \textbf{\bibinfo{volume}{90}}, \bibinfo{pages}{235205}
  (\bibinfo{year}{2014}).

\bibitem{Jones_1989}
\bibinfo{author}{Jones, R.~O.} \& \bibinfo{author}{Gunnarsson, O.}
\newblock \bibinfo{title}{The density functional formalism, its applications
  and prospects}.
\newblock \emph{\bibinfo{journal}{Reviews of Modern Physics}}
  \textbf{\bibinfo{volume}{61}}, \bibinfo{pages}{689} (\bibinfo{year}{1989}).

\bibitem{Hellman_2004}
\bibinfo{author}{Hellman, A.}, \bibinfo{author}{Razaznejad, B.} \&
  \bibinfo{author}{Lundqvist, B.~I.}
\newblock \bibinfo{title}{Potential-energy surfaces for excited states in
  extended systems}.
\newblock \emph{\bibinfo{journal}{The Journal of chemical physics}}
  \textbf{\bibinfo{volume}{120}}, \bibinfo{pages}{4593--4602}
  (\bibinfo{year}{2004}).

\bibitem{Kowalczyk_2011}
\bibinfo{author}{Kowalczyk, T.}, \bibinfo{author}{Yost, S.~R.} \&
  \bibinfo{author}{Voorhis, T.~V.}
\newblock \bibinfo{title}{Assessment of the $\delta${SCF} density functional
  theory approach for electronic excitations in organic dyes}.
\newblock \emph{\bibinfo{journal}{The Journal of Chemical Physics}}
  \textbf{\bibinfo{volume}{134}}, \bibinfo{pages}{054128}
  (\bibinfo{year}{2011}).

\bibitem{Gali_2009}
\bibinfo{author}{Gali, A.}, \bibinfo{author}{Janz{\'e}n, E.},
  \bibinfo{author}{De{\'a}k, P.}, \bibinfo{author}{Kresse, G.} \&
  \bibinfo{author}{Kaxiras, E.}
\newblock \bibinfo{title}{Theory of spin-conserving excitation of the {N}- {V}-
  center in diamond}.
\newblock \emph{\bibinfo{journal}{Physical review letters}}
  \textbf{\bibinfo{volume}{103}}, \bibinfo{pages}{186404}
  (\bibinfo{year}{2009}).

\bibitem{Davidsson_2018}
\bibinfo{author}{Davidsson, J.} \emph{et~al.}
\newblock \bibinfo{title}{First principles predictions of magneto-optical data
  for semiconductor point defect identification: the case of divacancy defects
  in 4{H}--{SiC}}.
\newblock \emph{\bibinfo{journal}{New Journal of Physics}}
  \textbf{\bibinfo{volume}{20}}, \bibinfo{pages}{023035}
  (\bibinfo{year}{2018}).

\bibitem{Mackoit_2019}
\bibinfo{author}{Mackoit-Sinkevi\v{c}ien\.{e}, M.}, \bibinfo{author}{Maciaszek,
  M.}, \bibinfo{author}{Van~de Walle, C.~G.} \& \bibinfo{author}{Alkauskas, A.}
\newblock \bibinfo{title}{{Carbon dimer defect as a source of the 4.1 eV
  luminescence in hexagonal boron nitride}}.
\newblock \emph{\bibinfo{journal}{Applied Physics Letters}}
  \textbf{\bibinfo{volume}{115}}, \bibinfo{pages}{212101}
  (\bibinfo{year}{2019}).

\bibitem{Hashemi_2021}
\bibinfo{author}{Hashemi, A.} \emph{et~al.}
\newblock \bibinfo{title}{Photoluminescence line shapes for color centers in
  silicon carbide from density functional theory calculations}.
\newblock \emph{\bibinfo{journal}{Physical Review B}}
  \textbf{\bibinfo{volume}{103}}, \bibinfo{pages}{125203}
  (\bibinfo{year}{2021}).

\bibitem{ham1968}
\bibinfo{author}{Ham, F.~S.}
\newblock \bibinfo{title}{Effect of {{Linear Jahn-Teller Coupling}} on
  {{Paramagnetic Resonance}} in a {{E}} 2 {{State}}}.
\newblock \emph{\bibinfo{journal}{Physical Review}}
  \textbf{\bibinfo{volume}{166}}, \bibinfo{pages}{307--321}
  (\bibinfo{year}{1968}).

\bibitem{Togo_2015}
\bibinfo{author}{Togo, A.} \& \bibinfo{author}{Tanaka, I.}
\newblock \bibinfo{title}{First principles phonon calculations in materials
  science}.
\newblock \emph{\bibinfo{journal}{Scripta Materialia}}
  \textbf{\bibinfo{volume}{108}}, \bibinfo{pages}{1--5} (\bibinfo{year}{2015}).

\bibitem{Silkinis_2024b}
\bibinfo{author}{Silkinis, R.} \emph{et~al.}
\newblock \bibinfo{title}{Optical lineshapes of the {C}-center in silicon from
  ab initio calculations: {I}nterplay of localized modes and bulk phonons}
  (\bibinfo{year}{2024}).
\newblock \urlprefix\url{https://arxiv.org/abs/2410.22158}.
\newblock \eprint{2410.22158}.

\end{thebibliography}


\begin{thebibliography}{1}
\expandafter\ifx\csname url\endcsname\relax
  \def\url#1{\texttt{#1}}\fi
\expandafter\ifx\csname urlprefix\endcsname\relax\def\urlprefix{URL }\fi
\providecommand{\bibinfo}[2]{#2}
\providecommand{\eprint}[2][]{\url{#2}}

\bibitem{Falk_2013}
\bibinfo{author}{Falk, A.~L.} \emph{et~al.}
\newblock \bibinfo{title}{Polytype control of spin qubits in silicon carbide}.
\newblock \emph{\bibinfo{journal}{Nature communications}}
  \textbf{\bibinfo{volume}{4}}, \bibinfo{pages}{1819} (\bibinfo{year}{2013}).

\bibitem{Davidsson_2018}
\bibinfo{author}{Davidsson, J.} \emph{et~al.}
\newblock \bibinfo{title}{First principles predictions of magneto-optical data
  for semiconductor point defect identification: the case of divacancy defects
  in 4h--sic}.
\newblock \emph{\bibinfo{journal}{New Journal of Physics}}
  \textbf{\bibinfo{volume}{20}}, \bibinfo{pages}{023035}
  (\bibinfo{year}{2018}).

\bibitem{Jin_2021}
\bibinfo{author}{Jin, Y.} \emph{et~al.}
\newblock \bibinfo{title}{Photoluminescence spectra of point defects in
  semiconductors: Validation of first-principles calculations}.
\newblock \emph{\bibinfo{journal}{Physical Review Materials}}
  \textbf{\bibinfo{volume}{5}}, \bibinfo{pages}{084603} (\bibinfo{year}{2021}).

\bibitem{Gordon_2015}
\bibinfo{author}{Gordon, L.}, \bibinfo{author}{Janotti, A.} \&
  \bibinfo{author}{Van~de Walle, C.~G.}
\newblock \bibinfo{title}{Defects as qubits in $3c\text{\ensuremath{-}}$ and
  $4h\text{\ensuremath{-}}\mathrm{SiC}$}.
\newblock \emph{\bibinfo{journal}{Physical Review B}}
  \textbf{\bibinfo{volume}{92}}, \bibinfo{pages}{045208}
  (\bibinfo{year}{2015}).

\bibitem{Falk_2014}
\bibinfo{author}{Falk, A.~L.} \emph{et~al.}
\newblock \bibinfo{title}{Electrically and mechanically tunable electron spins
  in silicon carbide color centers}.
\newblock \emph{\bibinfo{journal}{Physical Review Letters}}
  \textbf{\bibinfo{volume}{112}}, \bibinfo{pages}{187601}
  (\bibinfo{year}{2014}).

\bibitem{Seo_2017}
\bibinfo{author}{Seo, H.}, \bibinfo{author}{Ma, H.}, \bibinfo{author}{Govoni,
  M.} \& \bibinfo{author}{Galli, G.}
\newblock \bibinfo{title}{Designing defect-based qubit candidates in wide-gap
  binary semiconductors for solid-state quantum technologies}.
\newblock \emph{\bibinfo{journal}{Physical Review Materials}}
  \textbf{\bibinfo{volume}{1}}, \bibinfo{pages}{075002} (\bibinfo{year}{2017}).

\bibitem{Razinkovas_2021}
\bibinfo{author}{Razinkovas, L.}, \bibinfo{author}{Doherty, M.~W.},
  \bibinfo{author}{Manson, N.~B.}, \bibinfo{author}{Van~de Walle, C.~G.} \&
  \bibinfo{author}{Alkauskas, A.}
\newblock \bibinfo{title}{Vibrational and vibronic structure of isolated point
  defects: The nitrogen-vacancy center in diamond}.
\newblock \emph{\bibinfo{journal}{Physical Review B}}
  \textbf{\bibinfo{volume}{104}}, \bibinfo{pages}{045303}
  (\bibinfo{year}{2021}).

\bibitem{Razinkovas_thesis}
\bibinfo{author}{Razinkovas, L.}
\newblock \emph{\bibinfo{title}{Vibrational properties and photoionization of
  color centers in diamond: theory and ab initio calculations}}.
\newblock Ph.D. thesis, \bibinfo{school}{Center for Physical Sciences and
  Technology (FTMC)} (\bibinfo{year}{2021}).
\newblock
  \urlprefix\url{https://talpykla.elaba.lt/elaba-fedora/objects/elaba:114210427/datastreams/MAIN/content}.

\bibitem{Alkauskas_2014}
\bibinfo{author}{Alkauskas, A.}, \bibinfo{author}{Buckley, B.~B.},
  \bibinfo{author}{Awschalom, D.~D.} \& \bibinfo{author}{Van~de Walle, C.~G.}
\newblock \bibinfo{title}{First-principles theory of the luminescence lineshape
  for the triplet transition in diamond nv centres}.
\newblock \emph{\bibinfo{journal}{New Journal of Physics}}
  \textbf{\bibinfo{volume}{16}}, \bibinfo{pages}{073026}
  (\bibinfo{year}{2014}).

\end{thebibliography}

\end{document}


\title{\textit{Supplementary Information} \\
Theory of the divacancy in 4H-SiC: Impact of Jahn--Teller effect on optical properties}

\author{Vytautas Žalandauskas}%
\email{vytautas.zalandauskas@ftmc.lt}
\affiliation{Center for Physical Sciences and Technology (FTMC),
  Vilnius LT-10257, Lithuania}

\author{Rokas Silkinis}%
\affiliation{Center for Physical Sciences and Technology (FTMC),
  Vilnius LT-10257, Lithuania}

\author{Lasse Vines}
\affiliation{Department of Physics/Centre for Materials Science and Nanotechnology, University of Oslo, P.O. Box 1048, Blindern, Oslo N-0316, Norway}

\author{Lukas Razinkovas}%
\affiliation{Center for Physical Sciences and Technology (FTMC),
  Vilnius LT-10257, Lithuania}
\affiliation{Department of Physics/Centre for Materials Science and Nanotechnology, University of Oslo, P.O. Box 1048, Blindern, Oslo N-0316, Norway}

\author{Marianne Etzelmüller Bathen}
\email{m.e.bathen@fys.uio.no}
\affiliation{Department of Physics/Centre for Materials Science and Nanotechnology, University of Oslo, P.O. Box 1048, Blindern, Oslo N-0316, Norway}

\maketitle
\tableofcontents
\newpage

\section*{Supplementary Note 1: Electronic structure of basal divacancies}

\begin{figure}
  \centering
  \includegraphics[width=0.7\textwidth]{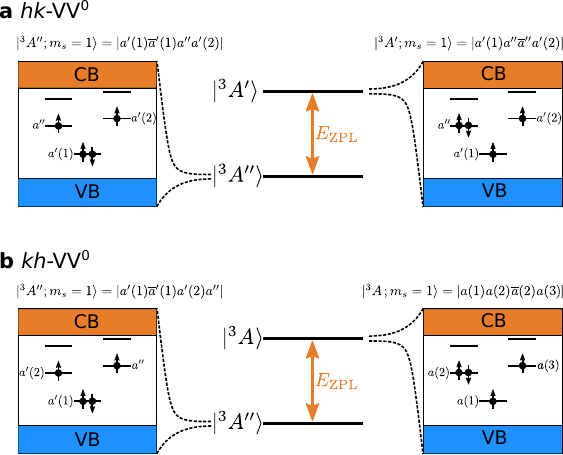}
  \caption{\textbf{Description of electronic structure of basal divacancies.} \textbf{a} Schematic representation of single-particle defect levels in the band gap of 4H-SiC showing the occupancy of these levels in the ground $^{3}A''$ state and excited $^{3}A'$ state of $hk$-VV$^{0}$. \textbf{b} Schematic representation of single-particle defect levels in the band gap of 4H-SiC showing the occupancy of these levels in the ground $^{3}A''$ state and excited $^{3}A$ state of $kh$-VV$^{0}$.}
  \label{fig:MO-basal}
\end{figure}

The electronic structure of neutral basal divacancies ($hk$-VV$^0$ and $kh$-VV$^0$) is presented in Supplementary Figure~\ref{fig:MO-basal}. The three lowest-lying molecular orbitals in the bandgap are derived from dangling carbon bonds. In the ground state these basal divacancies exhibit reduced $C_{1h}$ symmetry with molecular orbitals that can either have $a'$ or $a''$ irreducible representations. The $hk$ divacancy displays three lower-lying levels within the band gap, ordered from lowest to highest energy as $a'(1)$, $a''$, and $a'(2)$ respectively (see Supplementary Figure~\ref{fig:MO-basal}a). The orbital symmetries of the ground state $kh$ divacancy, arranged from lowest to highest energy, are denoted as $a'(1)$, $a'(2)$, and $a''$ while the excited state of the $kh$ divacancy has $C_{1}$ symmetry with $a$-symmetry orbitals (see Supplementary Figure~\ref{fig:MO-basal}b). For the $hk$ divacancy, the ground state $^{3}A''$ can be described by a single-determinant wave function $|a'(1)\bar{a}'(1)a''a'(2)|$ while the excited state will be $^{3}A'$ with a single-determinant wave function $|a'(1)a''\bar{a}''a'(2)|$. For the $kh$ divacancy the ground state has $A''$ orbital symmetry, designated as $^{3}A''$, whereas the excited triplet state with $A$ orbital symmetry is denoted as $^{3}A$.

\section*{Supplementary Note 2: Transition dipole moments}

Perpendicular ($\mu_{\perp}$) and parallel ($\mu_{\parallel}$) transition dipole moments (TDM), calculated using the r$^{2}$SCAN density functional, are presented in Supplementary Table~\ref{tab:TDM}. The transition dipole moments are estimated using the expression for molecular orbital states as
\begin{equation}
    \boldsymbol{\mu} = \left\langle \psi_i \right| q \, \boldsymbol{r} \left| \psi_j \right\rangle,
\end{equation}
where the Slater--Condon rule enables evaluation of the transition dipole moment for Kohn--Sham single-particle states, $\psi_i$ and $\psi_j$, rather than for many-electron states, $\Psi_i$ and $\Psi_j$. Here, $\psi_i$ and $\psi_j$ are computed using single-particle Kohn--Sham orbitals obtained from the triplet electronic configuration, a standard approach in density functional theory calculations.

\begin{table}
  \caption{\label{tab:TDM} Transition dipole moments (in units of Debye$^2$) for each divacancy configuration for absorption.}
  \begin{ruledtabular}
    \begin{tabular}{ccccc}
       & \textit{hh} & \textit{kk} & \textit{hk} & \textit{kh}\\
      \hline
      $\left| \overline{\mu}_{\perp} \right|^2$     & 27.16 & 25.81 & 33.15 &  6.32 \\
      $\left| \overline{\mu}_{\parallel} \right|^2$ &     0 &     0 &     0 & 45.56 \\
    \end{tabular}
  \end{ruledtabular}
\end{table}

\section*{Supplementary Note 3: Zero-phonon line energies and zero-field splitting}

Comparison of calculated and experimental zero-phonon line (ZPL) energies for four divacancy configurations are presented in Supplementary Table~\ref{tab:ZPL_SM}. Our ZPL values were computed using SCAN and r$^2$SCAN functionals. Previous theoretical ZPL energy predictions for available divacancy configurations in 4H-SiC with different density functionals are also presented in the table, along with brief computational details shown below. Similarly, Supplementary Table~\ref{tab:ZFS_SM} presents the zero-field splitting (ZFS) values in the same manner. From the mean absolute errors, we can observe that SCAN and r$^2$SCAN functionals enhance the accuracy compared to other calculations and come closest to the experimental values while requiring only a fraction of the computational cost needed for hybrid functionals.

\begin{table}
\centering
\caption{Calculated and experimental zero-phonon line (ZPL) energies (in units of eV) for spin-conserving transitions computed using different levels of theory for the neutral divacancy centers in 4H-SiC. Previous theoretical predictions on ZPL energies are also shown. MAE denotes the mean absolute error (in units of eV).}
\label{tab:ZPL_SM}
\begin{ruledtabular}
\begin{tabular}{ccccccc}
 &  & $hh$ & $kk$ & $hk$ & $kh$ & MAE \\
\hline
\multirow{2}{*}{This work} & SCAN & 1.129 & 1.123 & 1.141 & 1.103 & 0.021 \\
& r$^2$SCAN & 1.079 & 1.081 & 1.100 & 1.062 & 0.035 \\
\cline{1-7}
\multirow{6}{*}{\parbox{2cm}{Previous theoretical work}}& \multirow{2}{*}{PBE} & 0.92$^{\mathrm{a}}$ & 0.94$^{\mathrm{a}}$ & 0.97$^{\mathrm{a}}$ & 0.95$^{\mathrm{a}}$ & 0.170 \\
& & 0.937$^{\mathrm{b}}$ & 0.951$^{\mathrm{b}}$ & 0.979$^{\mathrm{b}}$ & $-$ & 0.158 \\
\cline{2-7}
& \multirow{3}{*}{HSE} & 1.056$^{\mathrm{a}}$ & 1.044$^{\mathrm{a}}$ & 1.103$^{\mathrm{a}}$ & 1.081$^{\mathrm{a}}$ & 0.044 \\
& & 1.221$^{\mathrm{b}}$ & 1.218$^{\mathrm{b}}$ & 1.269$^{\mathrm{b}}$ & $-$ & 0.122 \\
& & 1.13$^{\mathrm{c}}$ & 1.14$^{\mathrm{c}}$ & 1.21$^{\mathrm{c}}$ & 1.24$^{\mathrm{c}}$ & 0.065 \\
\cline{2-7}
& DDH & 1.196$^{\mathrm{b}}$ & 1.201$^{\mathrm{b}}$ & 1.259$^{\mathrm{b}}$ & $-$ & 0.105 \\
\cline{1-7}
& Expt.~\cite{Falk_2013} & 1.095 & 1.096 & 1.150 & 1.119 & \\
\end{tabular}
\end{ruledtabular}
\begin{flushleft}
$^{\mathrm{a}}$Reference~\cite{Davidsson_2018}: Calculations were carried out using the VASP code with PAW pseudopotentials (PPs). The Brillouin zone was sampled with the $\Gamma$ point. For PBE calculations, a $(10 \times 10 \times 3)$ supercell was used, while a $(8 \times 8 \times 3)$ supercell was used for HSE calculations with PBE structure.\\
$^{\mathrm{b}}$Reference~\cite{Jin_2021}: Calculations were carried out using the Quantum Espresso package with ONCV PPs. The plane-wave energy cutoff was set to 80 Ry. The Brillouin zone was sampled with the $\Gamma$ point. The ZPL energy values are with finite-size corrections, which were calculated as the difference in ZPL energy values between the $(8 \times 8 \times 2)$ and the $(5 \times 5 \times 2)$ supercell at the PBE level of theory.\\
$^{\mathrm{c}}$Reference~\cite{Gordon_2015}: Calculations were carried out using the VASP code with PAW PPs. The plane-wave energy cutoff was set to 400 eV. The Brillouin zone was sampled using a $2 \times 2 \times 2$ k-point mesh. A $(4 \times 3 \times 1)$ supercell was used for the neutral divacancies.
\end{flushleft}
\end{table}

\begin{table}
\centering
\caption{Calculated and experimental zero-field splitting (ZFS) values (in units of GHz) computed using different levels of theory for the neutral divacancy centers in 4H-SiC. Previous theoretical predictions on ZFS values are also shown. MAE is the mean absolute error (in units of GHz).}
\label{tab:ZFS_SM}
\begin{ruledtabular}
\begin{tabular}{ccccccc}
 &  & $hh$ & $kk$ & $hk$ & $kh$ & MAE \\
\hline
\multirow{2}{*}{This work} & SCAN & 1.336 & 1.271 & 1.320 & 1.279 & 0.026 \\
& r$^2$SCAN & 1.335 & 1.285 & 1.328 & 1.292 & 0.024 \\
\cline{1-7}
\multirow{3}{*}{\parbox{2cm}{Previous theoretical work}}& \multirow{2}{*}{PBE} & 1.358$^{\mathrm{a}}$ & 1.321$^{\mathrm{a}}$ & 1.320$^{\mathrm{a}}$ & 1.376$^{\mathrm{a}}$ & 0.052 \\
& & $1.387 (1.682)^{\mathrm{b}}$ & $1.349 (1.635)^{\mathrm{b}}$ & $1.306 (1.580)^{\mathrm{b}}$ & $1.356 (1.641)^{\mathrm{b}}$ & $0.064 (0.335)$ \\
\cline{2-7}
& HSE & 1.329$^{\mathrm{c}}$ & 1.307$^{\mathrm{c}}$ & 1.363$^{\mathrm{c}}$ & 1.314$^{\mathrm{c}}$ & 0.033 \\
\cline{1-7}
& Expt.~\cite{Falk_2013} & 1.336 & 1.305 & 1.334 & 1.222 & \\
\end{tabular}
\end{ruledtabular}
\begin{flushleft}
$^{\mathrm{a}}$Reference~\cite{Falk_2014}: Calculations were carried out using the VASP code with PAW pseudopotentials (PPs). The plane-wave energy cutoff was set to 420 eV. The Brillouin zone was sampled with the $\Gamma$ point. The 4H-SiC slab was built up from 1200 atoms, about 40 \AA \ thick along the c-axis.\\
$^{\mathrm{b}}$Reference~\cite{Seo_2017}: Calculations were carried out using the Quantum Espresso package with ONCV PPs, while the values in parentheses are with PAW PPs. The plane-wave energy cutoff was set to 75 Ry. The Brillouin zone was sampled with the $\Gamma$ point. A 480-atom supercell was used.\\
$^{\mathrm{c}}$ Reference~\cite{Davidsson_2018}: Calculations were carried out using the VASP code with PAW PPs. The Brillouin zone was sampled with the $\Gamma$ point. A $(8 \times 8 \times 3)$ supercell was used for HSE calculations with PBE structure.
\end{flushleft}
\end{table}

\section*{Supplementary Note 4: Details on calculation of adiabatic potential energy surfaces}

To obtain the adiabatic potential energy surface (APES) of the lower branch of the $^{3}E$ state, depicted in Fig.~3a of the main text, we first calculated the potential energy curve (PEC) to obtain the $K$ (elastic force constant), $F$ (linear vibronic constant), and $G$ (quadratic vibronic constant) parameters. At first, two calculations were performed: one where the geometry was relaxed with the $a_{1}e_{x}^{1.5}e_{y}^{1.5}$ electronic configuration, maintaining the $C_{3v}$ symmetry, and another where the geometry was relaxed with the $a_{1}e_{x}^{2}e_{y}^{1}$ electronic configuration, allowing the symmetry to break and the system to relax to a Jahn--Teller minimum with $C_{1h}$ symmetry. We then obtained the projected displacements along the $e$-symmetry breaking direction between these two relaxed structures. We calculated the mass-weighted displacements $\Delta Q$ (amu$^{0.5}$\AA) between these two equilibrium geometries to be 0.519 (amu$^{0.5}$\AA) for the $hh$-VV$^{0}$ and 0.472 for $kk$-VV$^{0}$.

We generated new geometries to probe the potential energy surface along the symmetry-breaking direction by systematically displacing the atoms along the $e_x$ component of the $e$-symmetry-breaking mode. This was achieved by projecting the displacements onto the $e_x$ direction using the irreducible representation matrices of the $C_{3v}$ point group. The displacements were then scaled by a range of factors (from -1.5 to 1.5 in increments of 0.1) to create a series of geometries along the APES. The resulting structures were then used as inputs for further single-shot VASP calculations with $a_{1}e_{x}^{2}e_{y}^{1}$ electronic configuration to map out the potential energy curve. At the $C_{3v}$ geometry, the calculations did not converge due to electronic degeneracy. However, the results from other geometries were sufficient to fit the computed energy values using Equation 1 as outlined in the main text. We found the corresponding vibronic coupling constants to be $K = 0.641$ eV/\AA$^2$, $F = 0.266$ eV/\AA, and $G = 0.064$ eV/\AA$^2$ for the $hh$ divacancy. For the $kk$ divacancy, these parameters were $K = 0.701$ eV/\AA$^2$, $F = 0.272$ eV/\AA, and $G = 0.061$ eV/\AA$^2$.

\section*{Supplementary Note 5: Benchmarking the diagonalization of the Jahn--Teller Hamiltonian}

We studied the convergence of calculated absorption lineshapes for the optical transition from ${^3A_{2}}$ to ${^3E}$, focusing on the parameters that define the Jahn--Teller (JT) Hamiltonian basis. In particular, we examined the convergence of absorption lineshapes for axial divacancies $hh$-VV$^0$ and $kk$-VV$^0$, while varying the number of effective modes $N_{\mathrm{eff}}$ (see Supplementary Figure~\ref{fig:abs_N_eff_conv}) and the number of excited phonons $n_{\mathrm{tot}}$ (see Supplementary Figure~\ref{fig:abs_n_tot_conv}). The theoretical framework for vibronic broadening of optical lineshapes in $A \to E$ transitions and the analysis of multi-mode $E \otimes (e \oplus e \oplus \cdots)$ JT systems is detailed in Ref.~\cite{Razinkovas_2021,Razinkovas_thesis}. Our results show that the absorption lineshapes for axial divacancies in 4H-SiC converge when $N_{\mathrm{eff}} = 9$ and $n_{\mathrm{tot}} = 6$.

\begin{figure}
\includegraphics[width=0.95\textwidth]{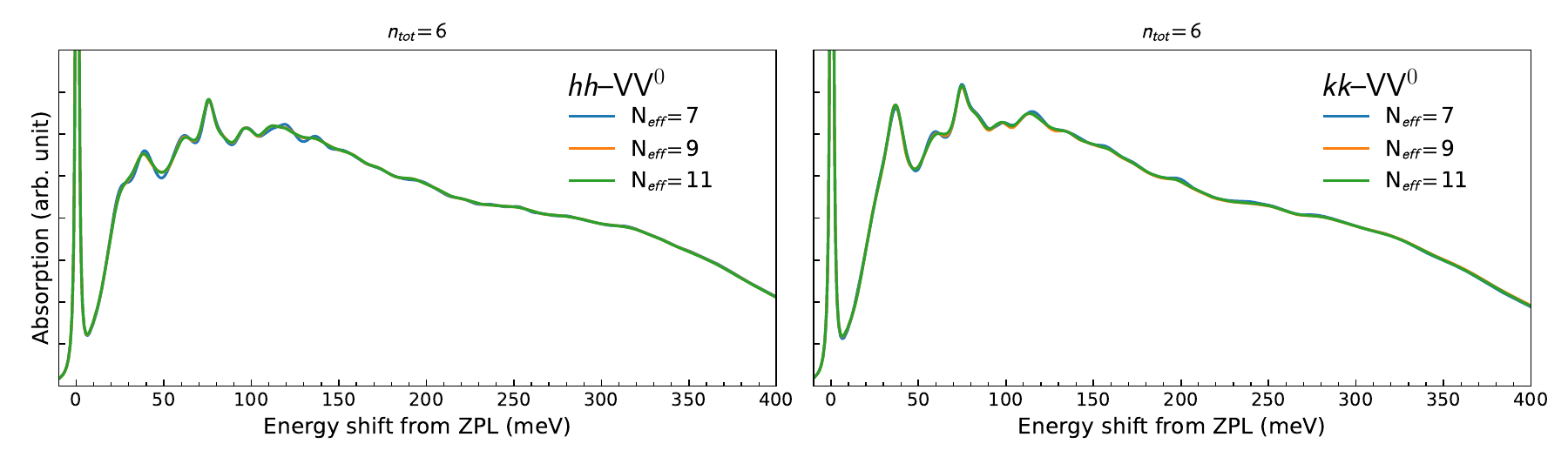}
\caption{\textbf{Convergence of calculated absorption lineshapes depending on the number of effective JT modes.} Convergence of the theoretical normalized luminescence lineshapes (in arbitrary units) for axial divacancies ($hh$-VV$^0$ and $kk$-VV$^0$) calculated using dynamical Jahn--Teller theory with respect to the number of effective modes $N_{\mathrm{eff}}$. The total number of excited phonons used for each lineshape was $n_{\mathrm{tot}}=6$.}
\label{fig:abs_N_eff_conv}
\end{figure}

\begin{figure}
\includegraphics[width=0.95\textwidth]{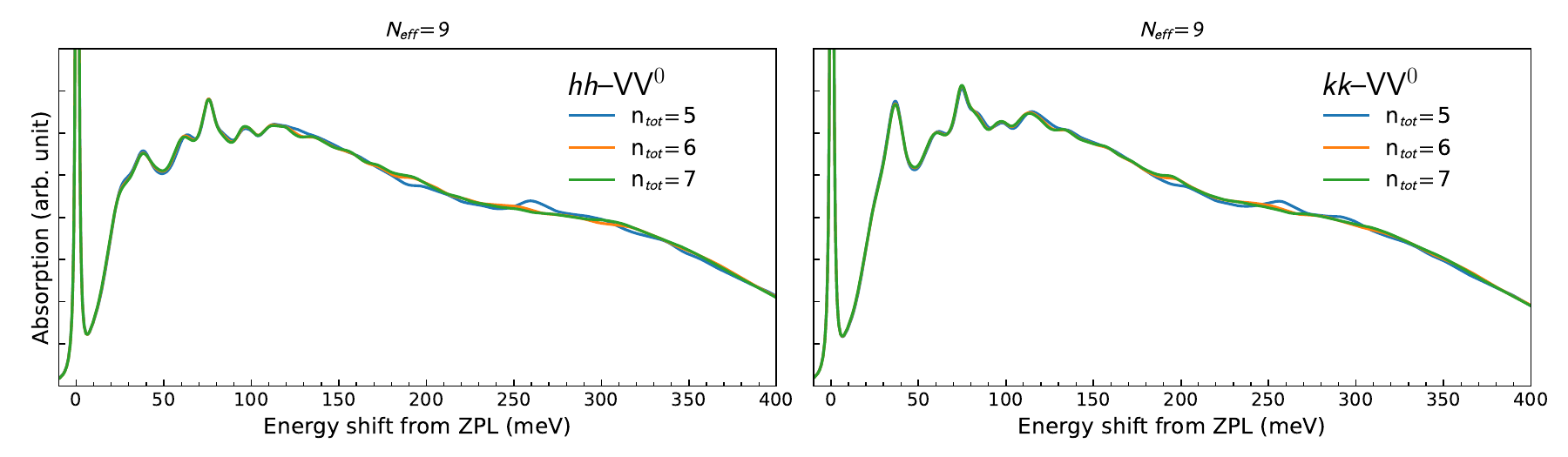}
\caption{\textbf{Convergence of calculated absorption lineshapes depending on the number of excited phonons.} Convergence of the theoretical normalized luminescence lineshapes (in arbitrary units) for axial divacancies ($hh$-VV$^0$ and $kk$-VV$^0$) calculated using dynamical Jahn--Teller theory with respect to the number of excited phonons $n_{\mathrm{tot}}$. The total number of effective modes used for each line shape was $N_{\mathrm{eff}}=9$.}
\label{fig:abs_n_tot_conv}
\end{figure}

\section*{Supplementary Note 6: Details of the embedding methodology}

The force constant embedding methodology outlined in Refs.~\cite{Alkauskas_2014,Razinkovas_2021} was used to examine the vibrational properties of large supercells. This method uses the short-range interatomic interactions in semiconductors to create a Hessian matrix for supercells with thousands of atoms. When constructing the Hessian matrix elements, the following criteria are used. If pairs of atoms are within a cutoff radius $r_{d}$ from any vacancy, matrix elements from the actual $(6\times6\times2)$ defect supercell are used. Otherwise, if two atoms are within another specified cutoff radius $r_{b}$, elements from the Hessian matrix of the bulk $(8\times8\times3)$ supercell are used, as shown in Supplementary Figure~\ref{fig:embedding}. In this study, we chose $r_{b} = 12.200~\text{\r{A}}$ and $r_{d} = 7.346~\text{\r{A}}$.

\begin{figure}[ht]
\includegraphics[width=0.70\textwidth]{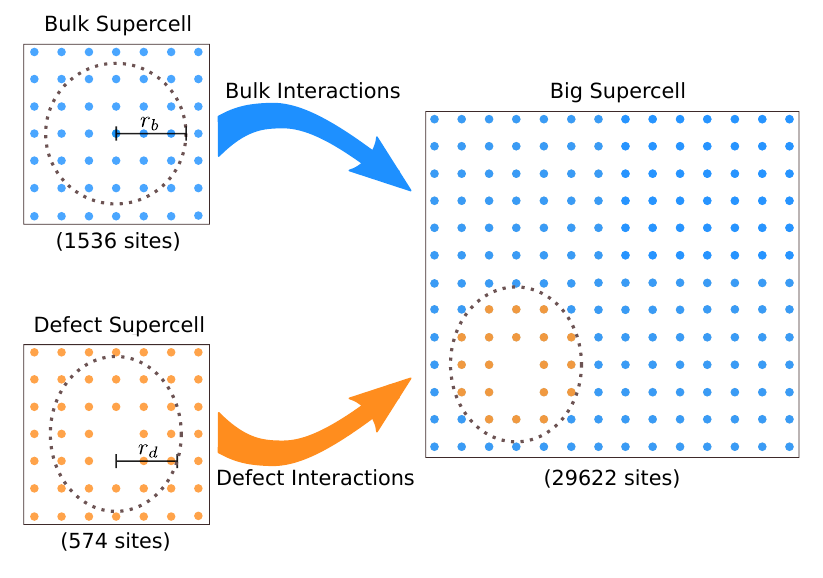}
\caption{\textbf{Embedding methodology.} Illustration of the force constant matrix embedding methodology for defect vibrational structure calculations.}
\label{fig:embedding}
\end{figure}

\bibliography{supplementary}